\documentclass[journal]{IEEEtran}

\usepackage{amsmath}
\usepackage{amsfonts}
\usepackage{amssymb}
\usepackage{amsthm}
\usepackage{tabularx}
\usepackage{mathrsfs} 
\usepackage{soul}

\usepackage{url}
\usepackage{hyperref}

\usepackage{stfloats}
\usepackage{float}
\usepackage{graphicx}
\usepackage{cite}
\usepackage{xcolor}
\usepackage{subfigure}

\usepackage{cleveref}

\usepackage{threeparttable}

\usepackage{algorithm}
\usepackage{algpseudocode}

\usepackage{pifont}

\usepackage{booktabs}

\usepackage{multirow}

\usepackage{rotating}

\usepackage{makecell}

\usepackage{array}

\makeatletter
\def\blfootnote{\xdef\@thefnmark{}\@footnotetext}
\makeatother

\usepackage{lipsum} 

\newtheorem{remark}{Remark}

\begin{document}

\title{\huge{Physical Layer Mutual Authentication in RIS-Aided Monostatic Backscatter Communications: \\ A Dual-Edged Analysis 
}} 
	\author{Masoud~Kaveh\IEEEmembership{}, Farshad Rostami Ghadi, \IEEEmembership{Member, IEEE}, Yishan Yang\IEEEmembership{}, Zheng Yan, \IEEEmembership{Fellow, IEEE}, and \\ Riku Jäntti, \IEEEmembership{Senior Member, IEEE}
	}
	\maketitle
	\vspace{-0pt}
 \begin{abstract}
Backscatter communication (BC) emerges as a pivotal technology for ultra-low-power energy harvesting applications, but its practical deployment is often hampered by notable security vulnerabilities. Physical layer authentication (PLA) offers a promising solution for securing BC by leveraging the unique characteristics of the communication medium. However, existing PLA approaches often fall short due to limited signal strength in practical BC scenarios and performance deterioration with increasing distance between the tag and the reader. Moreover, achieving mutual authentication has been largely neglected in current PLA schemes, given the passive nature of tags and their limited computational and energy resources.
This paper introduces a reconfigurable intelligent surfaces (RIS)-aided PLA scheme based on the physical features of received signals at legitimate endpoints through cascade links in monostatic BC (MBC) systems. By considering a RIS operating in its near-optimal conditions between a tag and a reader, the proposed PLA leverages the RIS-enhanced power delivery detected by the tag’s energy detector and the optimized received signal strength (RSS) at the reader’s signal processing unit, leading to address the conventional challenges of mutual authentication, low PLA performance, and limited secure coverage area inherent in BC systems.
Through theoretical analysis and extensive simulations, we show that as long as RIS is controlled by a trusted party in the network, it can boost the authentication performance across different system settings and strengthen the security features. 
Additionally, we conduct an analysis to explore the potential security threats when the RIS is compromised by an adversary by assessing its impact on the system’s PLA performance and secrecy capacity, providing a comprehensive understanding of the security implications for RIS-aided MBC under such circumstances.
	\end{abstract}
	\maketitle
 \blfootnote{\noindent  This work is supported in part by the Academy of Finland under Grants 345072 and 350464.}

	 	\blfootnote{\noindent M. Kaveh and R. Jäntti are with the Department of Information and Communication Engineering, Aalto University, Espoo, Finland. (e-mail: $\rm masoud.kaveh@aalto.fi, riku.jantti@aalto.fi$)}
	 	
	 	\blfootnote{\noindent  F. R. Ghadi is with the Department of Electronic and Electrical Engineering, University College London, WC1E
		6BT London, UK. (e-mail: $\rm f.rostamighadi@ucl.ac.uk$).}

   \blfootnote{\noindent Y. Yang and Z. Yan are with the School of Cyber Engineering, Xidian University, Xi'an, China, (e-mail: $\rm ysyangxd@stu.xidian.edu.cn, \\ zyan@xidian.edu.cn$)}

	


\section{Introduction}
\IEEEPARstart{B}{ackscatter} communication (BC) stands as an increasingly crucial wireless communication technology, especially valuable in low-power settings where data transfer rates are modest. This principle applies to domains such as the expansive Internet of Things (IoT) and radio frequency identification (RFID), where monostatic BC (MBC)’s capability to utilize radio frequency (RF) signals for both power and data transmission is revolutionary. Essentially, MBC enables numerous IoT devices to maintain connectivity while consuming minimal energy
\cite{BC_Battery_CST2023}.
However, MBC struggles with keeping communication strong and uninterrupted over distances, which is a real sticking point for large-scale networks. Security, too, is a significant concern as the passive nature of the backscatter devices (BDs) makes them vulnerable to a variety of security breaches. These issues underscore the urgent need for innovation, not just to improve the signals and extend the reach of these communications but to make sure they're secure, paving the way for a reliable and safe IoT world \cite{BC_survey_01}.

Reconfigurable intelligent surfaces (RIS) has recently emerged as a pivotal supplement technology for BC \cite{RIS_BC_frontier_6G,RIS_BC_Survey_Proceeding}. Composed of numerous reflective passive elements, RIS exploits meta-materials' properties to dynamically manipulate RF signals, fostering improvements in wireless propagation environments \cite{ref18}. This leads not only to enhanced communication performance in BC but also to extended transmission ranges without additional power expenditure \cite{ref20}.
Moreover, the smart reflectivity characteristic of RIS significantly upscales the BC system's capacity, allowing for an uptick in data rates and overall throughput \cite{ref26}. RIS also aids in refining the signal at the receiver end, elevating the detection accuracy of the backscattered signals \cite{ref36}. 
Energy efficiency is another hallmark of RIS integration into BC by directing additional power links, thereby achieving higher performance gains with less transmit power \cite{ref29}. 
RIS also holds promise in energy harvesting domains; by focusing RF energy towards BDs, it supports more effective energy collection, pushing towards a sustainable communicative framework \cite{ref3941}. RIS's versatile application in various BC systems
demonstrates its role as a multifaceted enabler, whether serving as an auxiliary element or a dedicated BD \cite{RIS_BC_Survey_Proceeding}.

Additionally, since BC signals are inherently passive and broadcast openly, they are prone to unauthorized interception and surveillance by adversarial actions. The resource limitations of BDs, such as constrained computational capabilities and minimal memory, further inhibit their use of advanced security protocols and cryptographic techniques (see Section \ref{subsec:crypto}). 
Consequently, physical layer authentication (PLA) emerges as a strong candidate for securing BC systems \cite{PLA_survey_01}. PLA leverages the inherent characteristics of the wireless medium to verify the legitimacy of communication devices. By assessing unique identifiers inherent in signal properties such as RF fingerprints, received signal strength (RSS), or channel state information (CSI), PLA can effectively detect and prevent unauthorized access. This method circumvents the resource-intensive processes associated with conventional cryptographic security, making it particularly suitable for the resource-constrained BDs
\cite{RSS_Auth_01}.

\subsection{Motivations}
The practical deployment of BC faces several critical security challenges, making the development of robust authentication protocols crucial. 
While, PLA has emerged as a promising alternative for securing BC by leveraging the unique characteristics of the wireless medium, existing PLA approaches encounter notable limitations that hinder their practical effectiveness. 
One major issue with current PLA schemes is is their reliance on weak direct links in BC systems, which makes it difficult to extract reliable physical layer attributes necessary for effective authentication, especially in environments with significant interference and noise.
Another critical challenge is the substantial reduction in PLA performance as the distance between the tag and the reader increases. This limitation results in a significantly restricted secure coverage area, as BC’s limited transmission range cannot maintain strong and reliable signals over larger indoor distances. 
Furthermore, achieving mutual authentication remains a largely unresolved issue in BC. While it is crucial for readers to authenticate tags, it is equally important for tags to verify the legitimacy of readers \cite{APAuth}. The passive nature of tags, coupled with their limited computational capabilities, complicates this process, leading to a gap in existing mutual authentication protocols for BC systems (see Section \ref{subsec:BC_PLA}).

Recent advancements in RIS-aided BC systems present a novel approach to addressing these challenges. RIS technology can dynamically manipulate the wireless environment to enhance signal quality, extend coverage, and improve energy efficiency \cite{RIS_BC_frontier_6G,RIS_BC_Survey_Proceeding,ref18,ref20,ref26,ref36,ref29,ref3941}. This capability suggests that RIS could play a pivotal role in strengthening PLA for BC systems by improving the received signal-to-noise ratio (SNR) and providing more reliable physical layer attributes for authentication even for longer distances.
Despite the promising potential of RIS, its application in PLA for BC systems remains largely unexplored. Moreover, the impact of a compromised RIS on PLA performance has not been thoroughly investigated, leaving potential vulnerabilities due to unauthorized control over RIS unaddressed and the overall resilience of RIS-aided BC systems in question. Understanding these aspects is crucial to developing a robust and practical authentication framework for BC systems.
This research is driven by the aforementioned challenges, with the motivation to bridge the security gap by leveraging the untapped potential of RIS to enhance PLA performance in BC systems. 

\subsection{Contributions}
The main contributions of our work can be summarized as follows.
\begin{itemize}
\item This study introduces a novel RIS-aided PLA scheme specifically designed for MBC systems. Our approach capitalizes on the unique capabilities of RIS to address the common challenges associated with PLA performance in traditional BC environments. By integrating RIS, we ensure the availability of reliable physical layer attributes that are crucial for authentication at BC endpoints, even over extended distances. This innovation also eliminates the need for more complex receivers in BC systems, simplifying the authentication process by leveraging the additional links provided by RIS.

\item Another vital advancement facilitated by our scheme is the ability of the tag to authenticate the reader, thereby realizing mutual authentication in BC systems, a feat not yet achieved in current designs. This development is made possible through the employment of RIS to maintain optimal power delivery at the tag's built-in energy detector circuit. This method effectively circumvents the need for computational complexity by relying on the voltage output profile generated from the inherent charging and discharging behaviors of the tag's internal capacitors.

\item We provide a comprehensive security analysis by envisioning the attacker in three roles: as a malicious tag, as a fake reader, and as an adversary who can maliciously control the RIS. We demonstrate that as long as the RIS is trustworthy and operates near its optimal configuration, the system is robust against various attack vectors and ensures efficient mutual authentication in MBC systems. Our analysis also delves into the security implications when the RIS is potentially malicious, providing important insights regarding MBC's different security aspects under such circumstances.

\item By conducting extensive simulations under various system settings and threat scenarios, our results illustrate the significant improvements a trusted RIS can introduce to the PLA performance compared to the secnarios and related works without considering RIS. We also measure the impact of a compromised RIS on the secrecy rate and PLA performance within the MBC systems. 
\end{itemize}

\subsection{Organization}
The rest of this paper is organized as follows. Section \ref{sec_related_works} presents related works on different PLA approaches for BC. Section \ref{sec_sys} delves into the studied system and threat models. Section \ref{sec_RISAuth} elaborates on the proposed RIS-aided PLA. Section \ref{sec_analy} provides a comprehensive security analysis for different attack scenarios. Section \ref{sec_results} presents the simulation results, and Section \ref{sec_conclusion} draws the conclusion of this paper.

\section{Related Works} \label{sec_related_works}
\subsection{Cryptography-based Authentication Methods in BC} \label{subsec:crypto}
Initial authentication strategies for BC systems, especially within RFID contexts, predominantly leveraged cryptographic techniques. Early protocols, referred to as the ultra-lightweight RFID authentication protocols, were designed for passive tags that possess limited processing power and storage capabilities, focusing on efficiency. These protocols employed straightforward bitwise operations including AND, OR, XOR (Exclusive OR), modular addition, and cyclic shifts. 
Nonetheless, concerns about the actual security provided by these protocols persist, as they often rely on superficially convincing yet unsubstantiated arguments \cite{RFID_Auth_07}.
Building on this, some studies introduced authentication protocols using hash-based and Rabin public key-based approaches for RFID systems \cite{Server_indp_RFID,USI_RFID_Auth}. While these protocols were claimed to be secure against various attacks, critiques have highlighted significant security flaws within these approaches \cite{RFID_Auth_Analyst_01}. To achieve a higher security level in MBC environments, protocols based on elliptic curve cryptography (ECC) 
were proposed \cite{RFID_Auth_ECC_01,RFID_Auth_ECC_02,RFID_Auth_ECC_03}. However, these methods are not feasible in practical MBC settings due to their computational complexity. Additionally, physical unclonable function (PUF)-based authentication methods have been suggested to enhance physical security in RFID systems \cite{RFID_Auth_PUF_01,RFID_Auth_PUF_02}. Despite their potential, the reliance on hash functions, PUF operations, and fuzzy extractors renders them impractical for recourse-limited passive tags within MBC systems.

\subsection{PLA Methods in BC} \label{subsec:BC_PLA}
The exploration of PLA presents a great alternative for cryptographic authentication methods in BC by capitalizing on the unique physical characteristics of the backscattered signals for authentication purposes. Various methodologies have been employed to enhance security features in BC systems by deploying PLA through recent years. 
In \cite{BC_PLA_01}, the authors proposed a PLA for identifying ultra-high frequency (UHF) RFID tags, focusing on enhancing security and reliability in MBC systems by exploiting unique physical characteristics inherent to individual tags. 
GenePrint \cite{BC_PLA_02} offered a generic and accurate PLA method that could be applied universally across different RFID systems, aiming to improve both security and system efficiency.
The study in \cite{BC_PLA_03} addressed the vulnerabilities of MBC systems to identity attacks. It proposed a PLA solution to prevent such attacks, enhancing the overall integrity and security of RFID systems.
Wang et al \cite{BC_PLA_04} presented an approach towards developing replay-resilient RFID authentication mechanisms. The study focused on coupling of two tags and signal randomization to ensure resilience against tag counterfeiting, signal replay, compensation, and brute-force feature reply attacks. 
Danev et al \cite{BC_PLA_05} presented a comprehensive investigation into the PLA of RFID transponders. The authors proposed multiple techniques for extracting physical-layer fingerprints from RFID devices and demonstrated the accurate identification of these transponders based on their unique physical properties. 
The authors in \cite{BC_PLA_06} addressed the vulnerability of BDs to active attacks due to their minimalist design and low-power radio technologies. They introduced ShieldScatter, which utilized low-cost tags to intentionally create multi-path propagation signatures, enabling the construction of sensitive profiles to identify the source of signals and detect potential threats. 
Wang et al \cite{BCAuth} presented BCAuth, a multi-stage authentication and attack tracing scheme designed to enhance the security of BDs. BCAuth utilized the physical spatial information of BDs to enhance the PLA for both static and mobile BDs. The scheme involves initial authentication based on BD identity with position information registration, followed by preemptive authentication and re-authentication based on spatial correlation of backscattered signal source locations associated with the BD. 
Li et al \cite{PLA_AmBC_NOMA} proposed a PLA for ambient BC-aided non-orthogonal multiple access (NOMA) systems. Three PLA schemes were proposed based on the multiplexing form of authentication tags: PLA with shared authentication tag, space division multiplexing authentication tags, and time-division multiplexing authentication tags. The authors also analyzed the probability of false alarm and probability of detection considering channel estimation errors.
Yang et al \cite{BatAu} introduced BatAu, a batch authentication scheme for authenticating multiple BDs in smart home networks. BatAu utilized physical layer features in multiplexing signals for authenticating batch BDs.
Zhang et al \cite{Cross-Domain-PLA} proposed FedScatter, a lightweight cross-domain authentication scheme for securing BC systems. FedScatter constructs device identity signatures from passive signal features generated by the tags and employed a federated learning model to aggregate device identity information across domains while preserving data privacy. 

The passive and resource-constrained nature of BDs poses significant challenges in establishing mutual authentication within BC systems, an aspect that remains largely unaddressed in existing literature. Many of the proposed schemes depend on sophisticated signal analyzers or require the deployment of multiple antennas at the reader, complicating the system architecture and escalating costs. Furthermore, these schemes typically necessitate operation under higher SNRs ranges to achieve acceptable PLA performance, limiting their practical applicability. Additionally, the limited coverage of these approaches poses a critical drawback; as the distance between the tag and the reader increases, the efficacy of PLA dramatically declines due to the inherently weak nature of backscatter signals in direct links.

\subsection{RIS-Aided PLA Approaches} \label{subsec:RIS_PLA}

The authors in \cite{RIS-PLA_01} introduced a RIS-assisted PLA system that enhances access security by allowing the transmitter to manipulate channel fingerprints via the RIS’s ON-OFF states. The paper leverages RSS-based spoofing detection to derive statistical properties of PLA, providing proofs of concept through experiments that demonstrate performance improvements under different transmitter placements.
The work in \cite{RIS-PLA_02} proposed using intrinsic PHY-layer features of RIS systems, including channel gain and background noise, to build a robust cover signal for authentication. It also employs asymmetric cryptography to secure tagged signals during transmission and applies statistical methods to assess authentication accuracy, enhancing security against unauthorized access.
The authors in \cite{RIS-PLA_03} proposed a hybrid RIS-based PLA that functions both as a reflector and a receiver to authenticate signals. By analyzing the channel response from a legitimate transmitter, this system improves the ability to distinguish between legitimate and malicious transmissions, offering a method against impersonation attacks near the transmitter.
The research in \cite{RIS-PLA_04} developed a PLA by utilizing both direct and cascaded channel features in RIS-assisted IoT systems. It details the statistical analysis of authentication verification through second-order statistics, aiming to improve detection accuracy and mitigate false alarms in dynamic communication environments.
The authors in \cite{RIS-PLA_05} integrated public-key algorithms and PLA to ensure secure message exchanges in varying signal conditions in vehicular communications. They employed RIS to boost SNR in challenging scenarios, enhancing detection probabilities and ensuring robust defense against both passive and active attacks.

The aforementioned works primarily focus on conventional communication systems and do not address the specific challenges posed by BC. Moreover, most of these approaches rely on cryptographic primitives, which may not be practical for resource-constrained BDs. Additionally, none of the studies explore the impact of a malicious RIS in a RIS-aided PLA scenario, overlooking a critical aspect of security in such systems. 
Table \ref{table_1} presents the unique aspects of our paper compared to the related works.  

\begin{table} [t]
    \centering
    \begin{threeparttable}
    \caption{Comparison of Other Authentication Approaches with Our Work Based on Different Key Criteria} 
    \label{table_1}
    {
    \begin{tabular}{c|ccccccc}
    \hline \hline
      Works   & BDA & RA & MRC & ESCR & CFA & MA & CSAI \\
    \hline
        \cite{APAuth} & $\times$ & \checkmark & -- & $\times$ & \checkmark & $\times$ & -- \\   
        \hline
        \cite{RFID_Auth_ECC_01,RFID_Auth_ECC_02,RFID_Auth_ECC_03, RFID_Auth_PUF_01,RFID_Auth_PUF_02} & \checkmark & $\times$ & -- & $\times$ & $\times$ & $\times$ & $\times$  \\
        \hline
        \cite{BC_PLA_01, BC_PLA_02, BC_PLA_03} & \checkmark & $\times$ & -- & $\times$ & $\times$ & $\times$ & $\times$\\
    \hline
    \cite{BC_PLA_04, BC_PLA_05, BC_PLA_06} & \checkmark & $\times$ & -- & $\times$ & $\times$ & $\times$ & $\times$\\
    \hline
        \cite{BCAuth, Cross-Domain-PLA} & \checkmark & $\times$ & -- & $\times$ & $\otimes$ & $\otimes$ & \checkmark \\
    \hline
        \cite{PLA_AmBC_NOMA, BatAu} & \checkmark & $\times$ & -- & $\times$ & \checkmark & $\times$ & $\times$\\
        \hline
        \cite{RIS-PLA_01,RIS-PLA_02 ,RIS-PLA_03 ,RIS-PLA_04 }  & -- & -- & $\times$ & \checkmark & \checkmark & $\times$ & $\otimes$\\
        \hline
        \cite{RIS-PLA_05}  & -- & -- & $\times$ & \checkmark & $\times$ & $\times$ & $\otimes$\\
    \hline
        Ours & \checkmark & \checkmark & \checkmark & \checkmark & \checkmark & \checkmark & \checkmark  \\
    \hline
    \end{tabular}
    } 
    \begin{tablenotes}
        \item BDA: BD authentication, RA: Reader authentication, MRC: Malicious RIS consideration, ESCR: Effective secure coverage range, CFA: Cryptography-free approach, MA: Mutual authentication, CSAI: Complex signal analyzer independence, \checkmark: Item is supported, $\times$: Item is not supported, $\otimes$: Item is conditionally supported, --: Not applicable. 
    \end{tablenotes}
\end{threeparttable}
\end{table}

\section{RIS-Aided MBC Overview} \label{sec_sys}
In this section, we explain the system model, threat model, and security goals for the studied RIS-aided MBC system.

\subsection{System Model}
\begin{figure}[t]
    \centering    \includegraphics[width=0.396\textwidth]{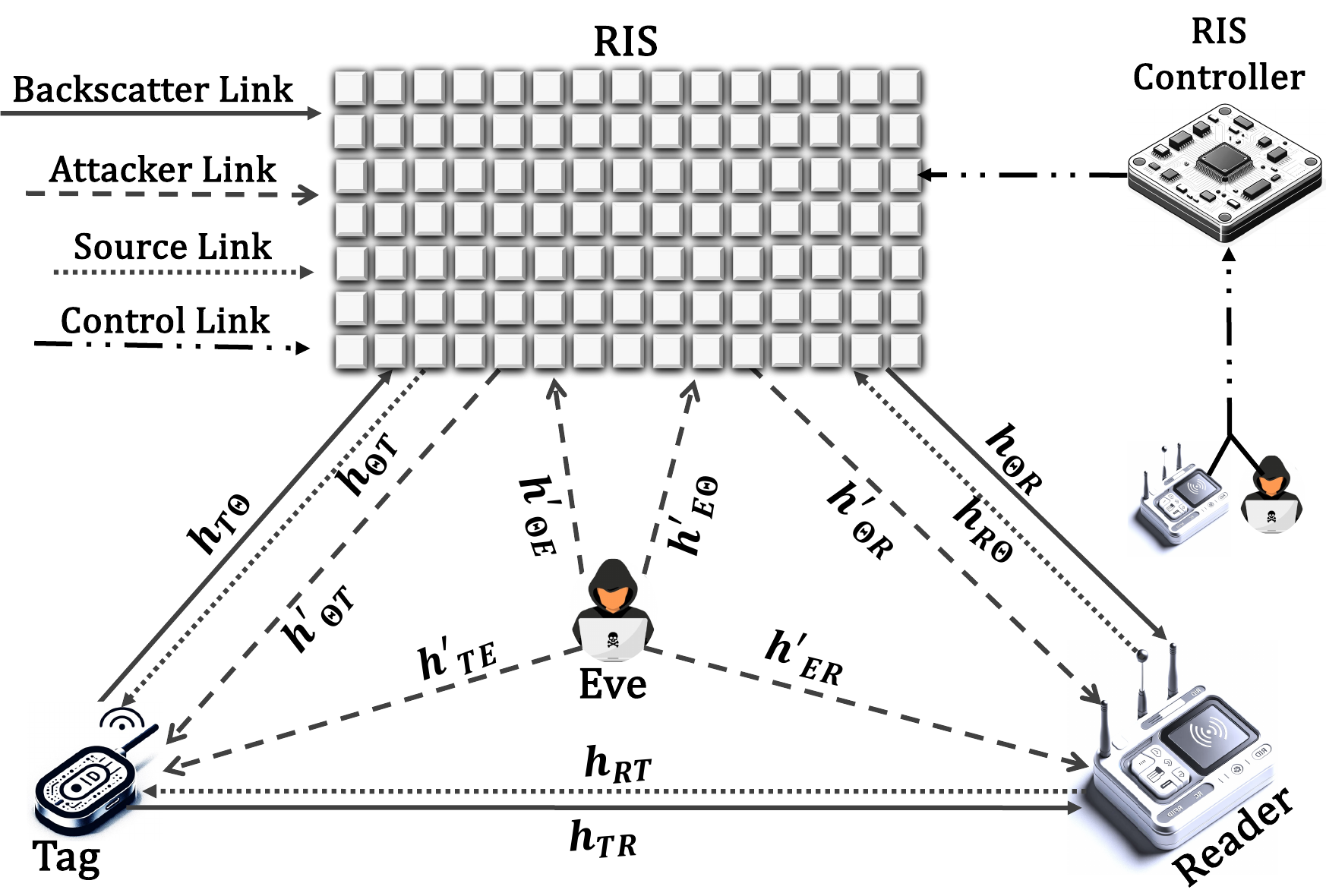}
    \caption{System model and security model depicting the tag, reader, and RIS configuration in RIS-aided MBC.}
    \label{fig:sysmodel}
\end{figure}

In our system model, we consider a MBC setup that includes a tag, a reader, an attacker (Eve), and a RIS as depicted in Fig.~\ref{fig:sysmodel}. 
We assume that the tag operates as a battery-less, single-antenna, passive devices, which harvests energy from a dedicated RF source's emitted signals through a simple energy detection circuit \cite{BD_Eng_detector1}. This simple yet effective mechanism allows the tag to modulate the incident waves from the reader with its data by reflecting the signal back to the reader in a time-division manner. 
The reader emits a continuous wave RF signal which powers the passive tag and is subsequently modulated and backscattered by the tag with its encoded data. The receiver circuitry with advanced signal processing capabilities within the reader is tasked with decoding this modulated backscatter signal and perform robust authentication process by extracting critical spatial information from the received signal.
The RIS is strategically positioned between the tag and reader to 1) direct the continuous waves from the reader towards tag to maximize the RF energy harvesting capability of the tag, thereby optimizing its battery charging process and ensuring sustained operation \cite{ref3941} and 2) to reflect the backscattered signal from the tag to the reader, thereby constructively reinforcing the communication pathway back to the reader, bolsters the RSS, and effectively extending the operational range and enhancing data throughput \cite{ref20,ref26}.
The RIS is assumed to be under the management of an advanced microcontroller that orchestrates the behavior of its constituent elements. The RIS comprises $N$ discrete reflecting elements, each capable of imposing an independent phase shift on the incident electromagnetic waves. This microcontroller-enabled RIS operates under a near-optimal condition protocol, which is defined by the intelligent adaptation of reflector states to maximize signal power at the receiver's end \cite{BjornsonCSI}.
%
Therefore, the received signal at the tag and the reader can be mathematically represented as \eqref{y_T_eq1} and \eqref{y_R_eq1}, respectively.
\begin{align} \label{y_T_eq1}
    y_T &= \sqrt{P_s} \left(h_{RT} + \mathbf{H}_{\Theta T}^{\mathcal{T}} \mathbf{\Phi}  \mathbf{H}_{R\Theta} \right) + n_T, \\ \label{y_R_eq1}
    y_R &= \sqrt{P_s} (h_{TR}h_{RT} + h_{TR}\mathbf{H}_{\Theta T}^{\mathcal{T}} \mathbf{\Phi}  \mathbf{H}_{R\Theta} + h_{RT}\mathbf{H}_{\Theta R}^{\mathcal{T}} \mathbf{\Phi}  \mathbf{H}_{T\Theta} \nonumber \\ &+ \mathbf{H}_{\Theta R}^{\mathcal{T}} \mathbf{\Phi}  \mathbf{H}_{T\Theta} \mathbf{H}_{\Theta T}^{\mathcal{T}} \mathbf{\Phi}  \mathbf{H}_{R\Theta} ) S(t)  + n_R,
\end{align}
where $P_s$ is the reader's power, $S(t)$ is the information signal backscattered from the tag with a unit power, $\mathbf{H}^{\mathcal{T}}$ is the transpose of matrix $\mathbf{H}$, $h_{RT}$, $h_{TR}$, $\mathbf{H}_{R\Theta}$, $\mathbf{H}_{\Theta R}$, $\mathbf{H}_{T\Theta}$, and $\mathbf{H}_{\Theta T}$ are reader-to-tag, tag-to-reader, reader-to-RIS, RIS-to-reader, tag-to-RIS, and RIS-to-tag  channel coefficients, respectively. 
$\mathbf{\Phi}$ represents the adjustable phase shift matrix of RIS for maximizing the received signal power at the tag and reader, which can be defined as $\mathbf{\Phi}=\text{diag}\left(\left[e^{j\theta_1}, e^{j\theta_2},...,e^{j\theta_N}\right]\right)$. Therefore, the RIS-aided channel coefficients like $\mathbf{H}_{T\Theta}$ and $\mathbf{H}_{R\Theta}$ contain the \emph{N} channel coefficients from the tag to the RIS and from reader to the RIS as $\mathbf{H}_{T\Theta}=d_{T\Theta}^{-\chi}.\left[h_{T{\Theta}_1}\mathrm{e}^{-j\alpha_1}, h_{T{\Theta}_2}\mathrm{e}^{-j\alpha_2},..., h_{T{\Theta}_N}\mathrm{e}^{-j\alpha_N}\right]$ and $\mathbf{H}_{R\Theta }=d_{R\Theta }^{-\chi}.\left[h_{{R\Theta }_1}\mathrm{e}^{-j\beta_{1}}, h_{{R\Theta }_2}\mathrm{e}^{-j\beta_{2}},..., h_{{R\Theta }_N}\mathrm{e}^{-j\beta_{N}}\right]$ (same will apply for $\mathbf{H}_{\Theta T}$ and $\mathbf{H}_{\Theta R}$), where $d_{T\Theta}$ and $d_{R\Theta}$ denote the distance between the tag and the RIS and the distance between the reader and RIS, respectively. 
Furthermore, the term $\chi$ indicates the path-loss exponent,
the terms $h_{T{\Theta}_n}$ and $h_{{R\Theta }_n}$ are the amplitudes of the corresponding channel coefficients, and the terms $\mathrm{e}^{-j\alpha_n}$ and $\mathrm{e}^{-j\beta_{n}}$ denote the phase of the respective links for $n\in\left\{1,2,...,  N\right\}$. In addition,
the terms $n_T$, $n_R$ show the additive white Gaussian noise (AWGN) at the tag and reader with zero mean and variances $\sigma^2_T$ and $\sigma^2_R$, respectively.
Given that the noise generated by the tag's antenna is significantly lower than the power of the signal received from the source \cite{ref57}, we will disregard it for the remainder of this paper.
As there may be obstructions impeding the direct link between the reader and the tag, we also assume that all links follow Raleigh fading distribution.

\subsection{Threat Model} \label{subsec:threat}

In practical MBC setups, distinctions are often made between readers and servers, with the latter typically viewed as a trusted entity managing various readers through secure channels by employing methods like OpenSSL. However, for the purposes of illustrating the effectiveness of the proposed PLA in leveraging RIS to secure communications between a reader and a tag, we simplify our model by not introducing additional entities. 
We treat the reader either as equivalent to a server or as being under secure server management, thus assuming it to be a trusted party within our framework. Furthermore, it is postulated that the reader has access to synchronized timing information, possibly derived from public GPS signals or a robust timekeeping system, the integrity of which is ensured by a timestamp validation algorithm. This synchronized timing is critical, as it allows the reader to send RF signals in a regimented manner
\cite{BCAuth}.

Equally important to our trust assumptions is the role of RIS. Within our PLA design, we first assume that RIS is predominantly considered as a trusted entity or being controlled by a trusted entity. Its microcontroller, governed by a trusted party such as a server or the reader itself, is presumed to be secure. In such a scenario, 
we assume that an adversary named Eve who has the means to intercept any exchanges between the reader and authorized tag in order to acquire the identity details of BDs. In summary, Eve is capable of executing the following attacks on the communication system:
\begin{itemize}
    \item \textit{Impersonation Attacks:} Eve attempts to impersonate the tag by sending signals that mimic the tag's profile. Eve can also appear as a fake reader, aiming at impersonating the reader from the tag's standpoint.  
    \item \textit{Man-in-the-Middle (MITM) Attacks:} Eve positions herself between the tag and reader to intercept and alter the communication.
    \item \textit{Replay Attack:} Eve captures a valid transmission and replays it later to masquerade as a legitimate party.
    \item \textit{Relay Attacks:} Eve captures the signal from the tag and relays it to the reader from a different location.
    \item \textit{Signal Jamming Attack:} Eve transmits a strong interfering signal to jam the communication between the tag and reader.
    \item \textit{Signal Injection Attacks:} Eve injects fabricated signals to mislead the reader into authenticating her as a legitimate endpoint.
\end{itemize}

While indoor RISs are predominantly contemplated for BC scenarios, making physical access challenging for compromising microcontroller \cite{Indoor_RIS_01, Indoor_RIS_02}, we yet consider a worst-case scenario where the RIS's microcontroller is manipulated or falls under the control of an adversary \cite{metasurfaceRISmanipu6G, RIS_attack01, RIS_attack02,RIS_attack03,RIS_attack05,RIS_attack06}. The various implications of such a compromise on system security and PLA performance are meticulously scrutinized and delineated as a separate concern within our security analysis section.

The security goals of this paper is to establish a robust RIS-aided PLA mechanism that effectively neutralizes all identified threats, ensuring secure communications between the tag and the reader in MBC systems. In addition, we delve into a thorough analysis to assess the potential security implications when the RIS is compromised by an adversary.

\section{The Proposed RIS-Aided PLA} \label{sec_RISAuth}
This section elaborates on the proposed RIS-aided PLA design for MBC, including comprehensive details of initialization and authentication phases. 

\subsection{Initialization Phase}

In practical implementations of the BC system, every BD needs to register with the server using its authentic identity.
In the studied system model in this paper, the reader initiates the process by transmitting power levels using a modulation method like on-off keying (OOK) to the tag. The RIS is employed to ensure optimal signal transmission towards the tag by optimally adjusting its elements' phase shift matrix, as it has been shown in \eqref{y_T_eq1}, during the initialization phase.
After receiving the RF signals from the reader through both direct and cascade links, the tag's energy detector circuit, which is typically includes a diode for rectification and a capacitor for storing the charge, being capable of capturing the incoming power levels from the reader  \cite{BD_Eng_detector1}, creates a voltage output profile, denoted as \( V^0_{out} \), based on the charging and discharging behavior of its internal capacitor. When the reader is transmitting (ON state), the circuit charges the capacitor, and when the reader is not transmitting (OFF state), the capacitor discharges. 
The charging and discharging behaviour of the tag's capacitor can be respectively shown as 
\begin{align}
    V_{out}(t) &= V_{peak} \cdot \left(1 - e^{-\frac{t}{\tau}}\right), \label{eq:charging}\\
    V_{out}(t) &= V_{out}(t_0) \cdot e^{-\frac{t}{\tau}}, \label{eq:discharging}
\end{align}
where \( V_{peak} \) is the peak voltage, corresponding to the maximum power level transmitted by the reader, \( \tau \) is the resistor–capacitor (RC) time constant of the circuit, with \( \tau = R \cdot C \),  and \( t_0 \) is the initial time at the start of the discharge phase. 
The relation between the incoming RF power and the output voltage $V_{out}$ is determined by the rectification efficiency of the diode, the charge storage capacity of the capacitor, and the discharge through the resistor $R$.
Then, the tag stores the \( V^0_{out} \) profile in its memory for later use during the authentication phase. Finally, the tag reflects its signal back to the reader through both direct and RIS-aided links, while the RIS is presumed to adjust the phase shifts of its elements to optimize the RSS at the reader.

\begin{figure} [t]
    \centering    \includegraphics[width=0.49\textwidth]{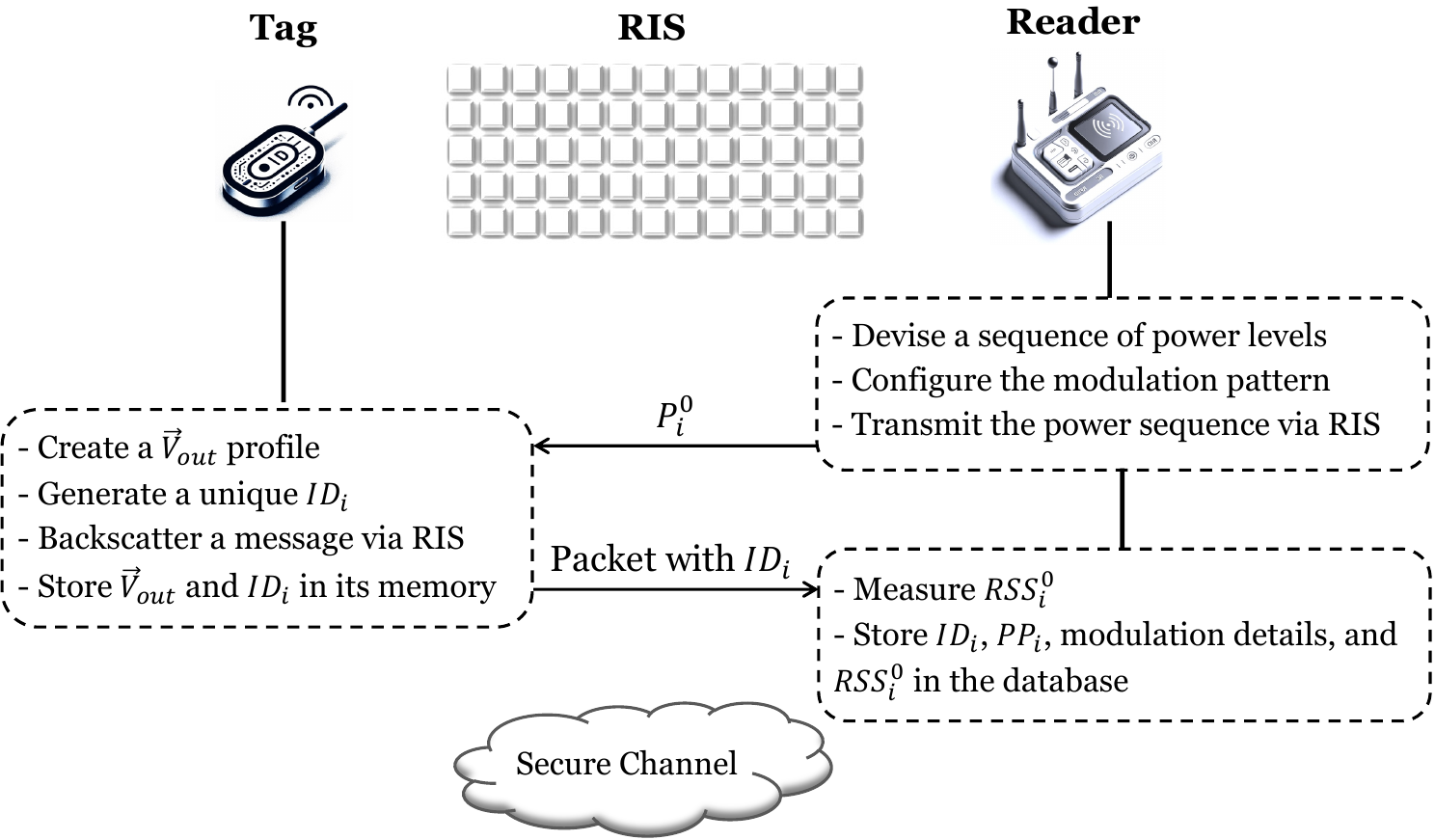}
    \caption{Initialization phase.}
    \label{fig:init}
\end{figure}

At the reader’s side, upon receiving the backscattered signal from the tag via both direct and RIS-assisted links, the reader calculates the RSS of the received signal as
\begin{align} \label{RSSeq}
     &RSS= P_s \ \times \nonumber \\ &{\left| \hspace{-2 pt}\left( \frac{h_{RT}}{d_{RT}^{\chi}} \right)^2 
     \hspace{-3 pt} + \hspace{-2 pt} \frac{2h_{RT}\sum_{n=1}^{N} h_{T\Theta_n} h_{R\Theta_n}}{d_{RT}^{\chi} d_{T\Theta}^{\chi} d_{R\Theta}^{\chi}} \hspace{-2 pt} + \hspace{-3 pt} \left( \hspace{-2 pt} \frac{\sum_{n=1}^{N} h_{T\Theta_n} h_{R\Theta_n}} {d_{T\Theta}^{\chi} d_{R\Theta}^{\chi}} \hspace{-2 pt} \right) ^2 \right|}^2,
\end{align}
where it is assumed that \( h_{xy} = h_{yx} \), \( x, y \in \{ R, T, \Theta \} \), due to channel reciprocity between the tag and the reader \cite{BCAuth}.
This RSS, named $\text{RSS}_i^0$, is then stored as a baseline reference for authenticating the $i$th tag in future communications alongside other information like the $i$th tag’s identification (ID) and its related modulation specifics and power profile (PP) in the corresponding row of the reader’s database for that tag as \{$\text{RSS}_i^0,\text{ID}_i, \text{PP}_i$\}. 
%
Fig. \ref{fig:init} depicts the different steps of the initialization phase of the proposed scheme for the \textit{i}th tag, which are executed through a secure channel \cite{RIS-PLA_02}. 

\subsection{Authentication Phase}

With this spatial information already stored in the tag and the reader, they can proactively authenticate each other during subsequent message transmissions. Following the protocols typical of BC systems, the reader first sends out a query command to select a tag and then sends a carrier signal that carries information from the tag through backscatter communications. 
Therefore, the reader begins the communication process by retrieving the corresponding power sequence and modulation details from the tag's row in its database, which contains the information established during the initialization phase, and then sending a sequence of power levels to charge the tag. The RIS is also employed to optimally deliver these power levels to the tag by adjusting the phase of its reflective elements \cite{ref3941}.

\begin{figure} [t]
    \centering    \includegraphics[width=0.49\textwidth]{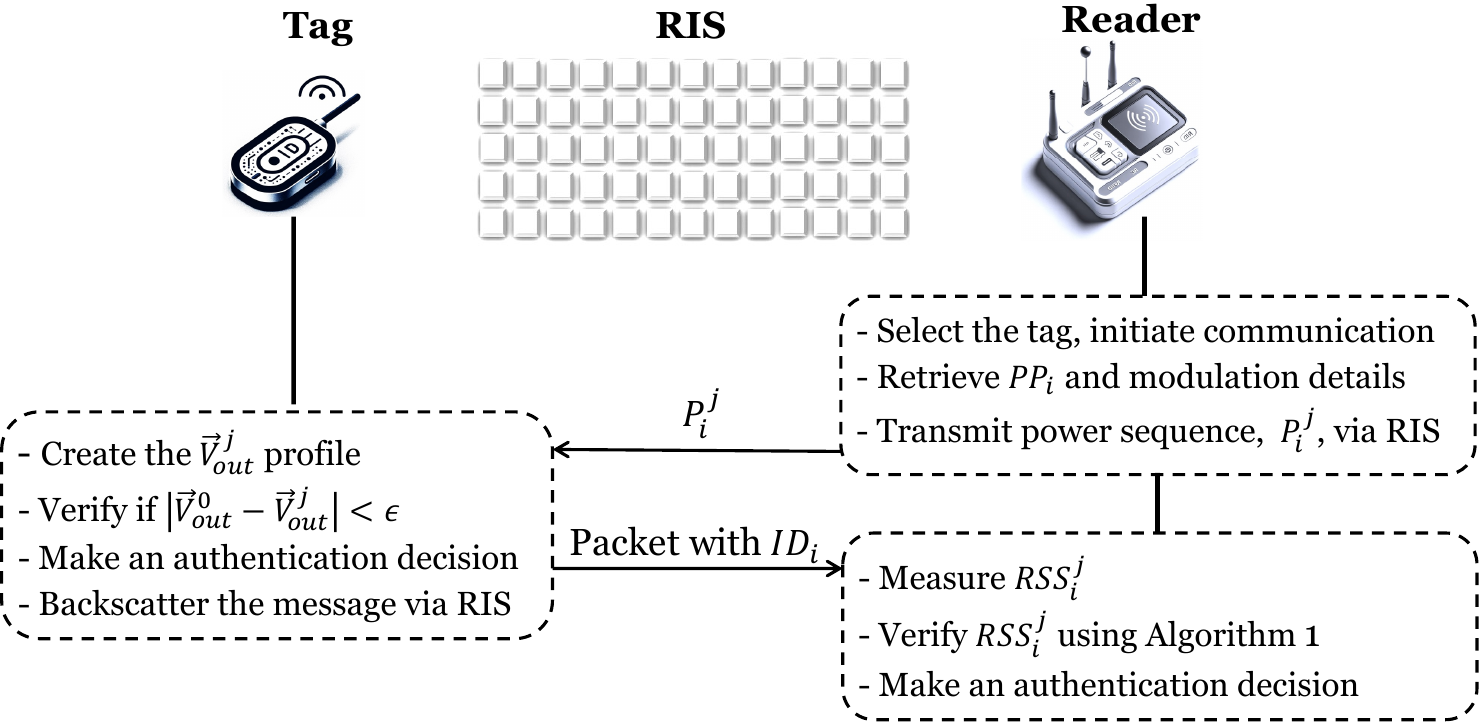}
    \caption{Authentication phase.}
    \label{fig:auth}
\end{figure}

Upon receiving power from the reader at $j$th transmission, the tag's energy detector measures the voltage output $V^j_{out}$, which reflects the charging and discharging of its internal capacitor. 
Then, the tag compares the observed $V^j_{out}$ profile against the stored profile; $V^0_{out}$, that includes the expected sequence of power levels received from the both direct and cascade links.
To do so, the tag sets predefined voltage thresholds on its comparator; a simple and low power circuit ideal for integration into energy-harvesting backscatter tags \cite{Tag-Comparator01}, to match the expected voltage levels resulting from the reader’s power levels. These thresholds are set based on the expected voltage levels that correspond to the received power states transmitted by the reader. 
When the detected voltage difference $\left|V^j_{out}-V^0_{out}\right| \leq \epsilon$, where $\epsilon$ is a predefined threshold, the reader is deemed authentic. Otherwise, the received signal is not considered valid.
\begin{remark}
    By optimally adjusting the phase of RIS's reflective elements, the reader can ensure the efficient transmission of its power levels, resulting in optimized harvested energy at the tag \cite{ref3941}. This process leads to the generation of a stable voltage profile at the tag’s energy detector, reflecting the consistent and reliable reception of power from the reader.
    The stability of the voltage profile is instrumental in the authentication mechanism employed by the tag since the tag compares the observed voltage output $V_{out}$ with the stored profile $V^0_{out}$ for authenticating the reader. This stable voltage profile, enabled by the RIS-assisted energy harvesting, enhances the accuracy and reliability of this comparison, allowing the tag to effectively discern between valid and invalid signals.
Additionally, a proper predefined voltage threshold, $\epsilon$, on the tag's comparator further reinforces the authentication process. 
\end{remark}
Once the tag authenticates the reader, it backscatters the signal, which includes its unique ID
and other information, in a message packet. The reader captures the backscattered signal (as shown in \eqref{y_R_eq1}) and begins the process of authenticating the tag. 
By dynamically adjusting its reflective elements, the RIS can create additional signal paths that constructively combine with the direct signal path, resulting in an enhancement of the received SNR at the reader \cite{RIS_MBC_01}. The phase configuration of the RIS, as represented by $\mathbf{\Phi}$, is crucial in ensuring that the reflected signals from the RIS are in phase with the direct signals from reader to tag and vice-versa, thus maximizing the received signal power at the tag and reader.
Therefore, the reader can compute the RSS of the backscattered signal as shown in \eqref{RSSeq}. 
Further, the reader verifies if the measured RSS matches the baseline measurements stored during the initialization phase, which corresponds to the tag's unique ID. In other words, the reader compares the current RSS measurement in \textit{j}th time slot, \{$\text{RSS}_i^j, \text{ID}_i$\}, with the registered one associated with the \textit{i}th tag in its database, \{$\text{RSS}_i^0, \text{ID}_i$\}.
It is worth mentioning that while the straightforward calculation of the RSS difference, $\mathcal{A} = |\text{RSS}_i^0 - \text{RSS}_i^j|$ may seem intuitive for comparing RSS values between different time slots, its effectiveness is limited in the context of RIS-aided BC systems. The presence of RIS introduces significant variations in the received signal power at the reader, leading to considerable differences in RSS values for different number of RIS elements. To address this issue and ensure a meaningful comparison of RSS distances for different values of $N$, we propose using Algorithm \ref{alg:calculate-ratio} for computing the ratio of RSS values between the baseline and current measurements. 
This algorithm accounts for the dynamic nature of the received signals by considering the relative magnitude of RSS values rather than their absolute difference. By normalizing the RSS comparison with respect to the baseline RSS, the algorithm facilitates a more robust and reliable authentication decision process. 
If the discrepancy between the measured RSS and the stored value in the $i$th tag’s profile in reader's database is within an acceptable range, the tag is authenticated, thereby confirming mutual authentication in the considered MBC system.
Discrepancies between the expected and measured value may prompt a re-authentication sequence or a security alert.
Fig. \ref{fig:auth} depicts the different steps of the mutual authentication phase of the proposed PLA for \textit{i}th tag in \textit{j}th time slot. 

\begin{algorithm}[t]
\caption{Calculate Ratio}
\label{alg:calculate-ratio}
\begin{algorithmic}[1]
\footnotesize
\State \textbf{Input:} Matrices $RSS_i^0$ and $RSS_i^j$
\State \textbf{Output:} Matrix $ratio$ containing the element-wise ratios of corresponding elements in $RSS_i^0$ and $RSS_i^j$
\State $[rows, cols] \gets \text{size}(RSS_i^j)$
\State $ratio \gets \text{zeros}(\text{size}(RSS_i^j))$
\For{$row = 1$ \textbf{to} $rows$}
    \For{$col = 1$ \textbf{to} $cols$}
        \If{$RSS_i^0[row, col] > RSS_i^j[row, col]$}
            \State $ratio[row, col] \gets \frac{RSS_i^j[row, col]}{RSS_i^0[row, col]}$
        \Else
            \State $ratio[row, col] \gets \frac{RSS_i^0[row, col]}{RSS_i^j[row, col]}$
        \EndIf
    \EndFor
\EndFor
\end{algorithmic}
\end{algorithm}

\section{Security Analysis and Discussion} \label{sec_analy}

In this section, we conduct a security analysis of the proposed PLA approach in the context of potential threats faced by RIS-aided MBC systems. These threats can be categorized into two main scenarios: 1) Trusted RIS Control: In this scenario, we assume the RIS is under the control of a trusted entity and operates in its near-optimal condition. Then, we examine two adversarial cases under such an assumption in which a malicious tag or a fake reader can execute the mentioned attack vectors outlined in Section \ref{subsec:threat}. 2) Compromised RIS: Here, we explore the consequences of a maliciously compromised RIS on PLA performance within an RIS-based MBC system. 

\subsection{Trusted RIS Control}

\subsubsection{Impersonation Attacks}
The integrity of MBC systems is significantly threatened by impersonation attacks, where an adversary attempt to masquerade as a legitimate tag or reader by replicating their signal profile. Considering a fake reader scenario, the attacker seeks to impersonate a genuine reader, potentially intending to unlawfully charge the tag’s battery and induce it to backscatter its signal even when not legitimately prompted to do so. 
The proposed PLA mechanism utilizes ability of RIS to ensure optimum power delivery at the tag \cite{ref3941} by intelligently adjusting its phase shift matrix. This meticulous adjustment maintains the integrity of the signal’s attributes harvested by the tag's energy detector's circuit, resulting in creating a robust output voltage profile and enabling the tag to accurately discern and authenticate the reader. Any impostor would need to precisely replicate the power sequence and the RIS-optimized signal trajectory, which the latter is a task nearly unattainable without direct control over the RIS configurations, which effectively thwarting unauthorized attempts at reader impersonation.
In the scenario when we have a malicious tag as Eve, our proposed PLA framework is inherently robust against such attacks by virtue of its dependence on RIS-assisted RSS at the reader. This parameter is challenging for Eve to precisely duplicate, as it is not only tied to the physical location of the legitimate tag but is also enhanced by the RIS to maintain a stable and robust profile, making any spoofing attempts resulting in a divergence between the observed and stored values.
This discrepancy is a direct consequence of the Eve’s inability to manipulate the RIS, ensuring that any unauthorized alteration to RSS is swiftly detected and thwarted, thereby upholding the integrity and security of the authentication process.

\subsubsection{MITM Attack}
In an MITM attack scenario within a RIS-enhanced MBC system, Eve strategically situates herself between the tag and the reader, aiming to surreptitiously intercept and potentially manipulate the signal exchanged between the legitimate communication endpoints. However, the presence of RIS-enhanced signals, which channel optimized beams directly towards the authorized reader, mitigates the impact of MITM attacks. Eve faces a formidable challenge in attempting to circumvent authentication by replicating the required RSS profile accurately. This entails the daunting task of precisely emulating the RIS-configured signal profile to maximize the received SNR at the reader.
Since it is assumed that Eve does not have any control over RIS's microcontroller in this attack scenario, any deviation from the expected signal characteristics is readily discernible to the reader.
Thus, the robustness of the RIS-enhanced signal configuration acts as a deterrent against MITM attacks.

\subsubsection{Replay Attack}
In a replay attack scenario within RIS-aided MBC systems, Eve might capture a valid transmission from the tag and attempt to replay it later, posing as the tag to deceive the reader. However, the RIS's reflective properties, optimized to the spatial location of the legitimate reader, ensure that any replayed signals would have distinct RSS profile. Since Eve's physical location would differ from that of the tag, and more importantly, she cannot optimally direct her backscattered signals to the reader through the RIS, any replayed signal she transmits cannot match the unique signal characteristics shaped by the RIS and initially recorded by the reader during the initialization phase. The reader, designed to expect signals with specific RSS signature, is thus able to detect such anomalies when a signal is seemingly backscattered from an unexpected location, exposing the replay attack. 

\subsubsection{Relay Attack}
In the case of relay attacks, Eve seeks to intercept the communication by capturing the signal transmitted by the tag and then relaying it to the reader from a different location. Our PLA scheme is informed by unique signal characteristics that arise from the specific placement and orientation of the tag and RIS. These characteristics are inherently difficult to replicate outside the original setting, making the task of a potential relay attacker significantly challenging, even when the attack is executed instantaneously.
When Eve captures and relays the backscattered signal, the characteristics of the transmitted signal will inherently differ due to the altered propagation path through RIS. In other words, the RIS configurations, which are tailored to optimize the RSS at the reader, would not align with those of a relayed signal. Since the reader has pre-stored the expected values of RSS from the initialization phase, it can therefore detect any anomalies in the relayed signal characteristics. 

\subsubsection{Signal Jamming Attack}
For the signal jamming attack scenario, Eve undertakes an approach by transmitting an interfering signal to disrupt the communication channel between the tag and the reader. In the proposed RIS-aided PLA, RIS can manipulate the signal propagation environment to bolster the strength of the legitimate signal or to negate the effects of jamming signals \cite{RISAntiJamm1}. 
This capability stems from the RIS's ability to vary the phase shifts of its elements in real-time, allowing it to craft constructive interference that amplifies the legitimate signal while concurrently creating destructive interference that weakens the jamming signal \cite{RISAntiJamm2}.
When Eve launches her jamming attempt, the RIS swiftly alters its configuration, concentrating the legitimate signal's energy towards the reader and dispersing or deflecting the energy of the jamming signal. This dynamic adjustment is possible because each RIS element can independently modify its phase shift, enabling the formation of a highly directional beam aimed at the reader \cite{RISAntiJamm3}. This mechanism significantly mitigates the impact of the jamming attack, preserving the integrity of the communication.

\subsubsection{Signal Injection Attack}
In the context of signal injection attacks, Eve attempts to deceive the reader by injecting fabricated signals, posing as a legitimate tag. The security strategy deployed in the proposed PLA shares similarities with the defenses against impersonation attack. 
In this scenario, the RIS, which is optimized based on the CSI between the legitimate parties, and controlled by the a trusted party, dynamically adjusts its reflective elements to manipulate the propagation environment to increase the SNR at the reader. 
By doing so, it ensures that signals follow precise trajectories tailored to legitimate tag transmissions.
Since Eve lacks control over the RIS, any injected signals introduced by her undergo modifications imposed by the RIS, which lead to deviations from the expected trajectories associated with genuine tag transmissions. As a result, when the reader measures the RSS of these injected signals, it detects inconsistencies compared to the baseline measurements stored during the initialization phase, and the illegitimate injected signal will be detected.

\subsection{Compromised RIS Control}
In this scenario, an attacker, through sophisticated means, might gain unauthorized access to the RIS controller. This breach could allow the attacker to manipulate the phase configurations, deliberately altering the signal paths and potentially weakening or even nullifying the secure zones intended for protected communication. Such control could enable the attacker to eavesdrop or disrupt communication by focusing or scattering the RF beams, hence compromising the confidentiality and reliability of the communication system. The very feature that allows the RIS to enhance connectivity i.e., its ability to shape and direct signals, can be exploited, turning it into a tool for signal interception or degradation \cite{metasurfaceRISmanipu6G}. In this subsection, we investigate the effect of a malicious RIS on PLA performance in MBC systems from different standpoints. 

\subsubsection{Jamming Attack} 
While RIS technology has been touted as an effective countermeasure against jamming attacks \cite{RISAntiJamm1,RISAntiJamm2,RISAntiJamm3}, recent insights reveal that a compromised RIS can also enhance jamming attacks against legitimate users \cite{RISJamm1,RISJamm2,RISJamm3,RISJamm4,RISJamm6}. Such manipulation can not only escalate the efficiency of the attack but also enable it to be conducted with minimal or no power expenditure on the part of the attacker, termed as \textit{green jamming} \cite{RISJamm3}. This makes the attack stealthier and more challenging to identify. Furthermore, the attacker can execute selective jamming by directing the interference precisely when the legitimate transmitter is active and subsequently redirecting it to eavesdrop, thereby gaining control over the communication timeline.
In our study, we posit that Eve has control over the RIS and manipulates its phase shifts to reflect signals from the tag to the reader in a manner that causes destructive interference between the direct and RIS-reflected links, thereby reducing the SNR at the reader. Under this assumption, two scenarios arise \cite{Mal_RIS_Dest_Beam}: a) if Eve is privy to the CSI of the tag-to-RIS and RIS-to-reader channels, she can optimize the signal's power at the tag via appropriate RIS phase adjustments. This allows her to pass the tag's authentication process, subsequently using the backscattered signal from the tag to jam the reader. b) Conversely, if Eve lacks knowledge of the CSI of tag-to-RIS and RIS-to-reader channels, she is unable to optimize the received power at the tag due to her inability to adjust the RIS phase optimally. 
Consequently, she might fail the tag's authentication test in the proposed PLA, preventing any backscattering from the tag and thus averting a jamming attack on the reader. This protection is thanks to the tag's authentication step outlined in our paper. However, in common BC systems that do not include reader authentication by the tag, she could still conduct a nearly successful jamming attack on the reader by (even) randomly adjusting the RIS phase shifts, without knowing the CSI between the Reader and RIS \cite{RISJamm4,RISJamm6}.

In another general scenario, Eve can launch an active jamming attack using RIS and her own signal power \cite{RISJamm2}, a method that doesn't rely on backscatter from the tag. This type of attack, though it requires Eve to consume her own energy, gains a significant advantage through the strategic use of the RIS. 
By controlling the RIS, Eve can optimize the path of her jamming signals towards the reader. This capability allows her to maximize the disruptive impact on the reader's received signal by ensuring that the jamming signal directly interferes with the communication from the tag. 
This targeted approach will still allow Eve to perform a highly efficient jamming attack with potentially less power than would be required without control over the RIS, which is difficult for traditional anti-jamming techniques to mitigate. 
Measuring the exact impact of malicious RIS-based jamming attacks on BC systems and identifying defensive strategies have been remained unexplored areas. Detailed exploration of these issues is beyond the scope of this paper and is identified as a promising direction for future research.

\subsubsection{Eavesdropping}

Eavesdropping in RIS-aided MBC systems presents a significant threat where Eve aims to intercept confidential communications between the tag and the reader (preferably without being detected). 
Although RIS has emerged as a promising tool to bolster the secrecy performance of wireless communication systems \cite{RIS_SG_PLS01}, in scenarios where the RIS is compromised, Eve can exploit the RIS's capability to focus and redirect signals to maximize the signal strength at her location, thus enhancing her ability to intercept these signals. This type of attack is particularly insidious because it can be executed with minimal detectable presence, leaving the legitimate users unaware of the ongoing breach \cite{Mal_RIS_sideband_Eve}.
Ideally, Eve would manipulate the electromagnetic wavefront to create dual-lobe reflections, fine-tuning the direction and power balance between herself and the reader to blend covertness with stealth. However, this advanced beam manipulation requires access to each metasurface element’s amplitude and phase settings, a challenging feat in practice due to the high-resolution phase shifters and amplitude controls required, which are typically not available in cost-effective metasurfaces \cite{Mal_RIS_sideband_Eve}.
Instead, an eavesdropping strategy that modifies the signal’s wavefront by altering the RIS’s bias lines can use periodic switching of the control lines to generate a sideband channel that mirrors the victim’s signal. This technique can precisely adjust these time-varying signals across the RIS's elements, directing the sideband towards Eve while keeping the primary signal path directed at the reader to remain undetected \cite{Mal_RIS_mmWave_Eve}.

The impact of a compromised RIS on the secrecy performance, particularly the secrecy rate, is profound \cite{metasurfaceRISmanipu6G,RIS_attack05}. In RIS-aided MBC systems, the secrecy rate could be defined as 
\begin{align} \label{SC1}
C_s(\gamma_R, \gamma_E)= {\Big[C_R-C_E\Big]}^+,
\end{align}
where $[X]^+=Max\{X,0\}$. $C_R=\log_2\left(1+\gamma_R\right)$ and $C_E=\log_2\left(1+\gamma_E\right)$ denote the wireless channel (including direct and RIS-aided links) capacity between the tag and reader, and the tag and Eve, respectively, in which $\gamma_R$ and $\gamma_E$ stand for the received SNR at the reader and Eve, respectively.
As can be seen in \eqref{SC1}, the secrecy rate heavily relies on the ability to maintain a high SNR differential between the reader and Eve. A compromised RIS disrupts this balance by enhancing Eve’s SNR while potentially reducing the reader’s SNR through destructive interference or signal redirection, leading to significantly degrading the secrecy performance.  
This effect is more pronounced as the number of RIS elements increases, providing Eve with greater flexibility to optimize her eavesdropping strategy (see Section \ref{Mal_RIS_SubSim} for more details). 
In the scenario when Eve has access to the CSI of both tag-to-RIS and RIS-to-reader channels, Eve can optimally configure the RIS to maximize the SNR at her location while effectively affecting the signal at the reader for the sake of attack covertness.
The enhanced SNR at Eve’s location directly translates to an increased data rate for her end, thus significantly lowering the system's secrecy rate. On the other hand, the reader’s ability to distinguish between authentic and intercepted signals diminishes, compromising the overall system security.
Without perfect CSI knowledge, Eve’s ability to precisely configure the RIS is limited. However, she can still attempt to optimize the signal paths towards her location by leveraging the known RIS-to-Eve channel information. Since Eve lacks the CSI of the tag-to-RIS and RIS-to-reader channels, she cannot optimally configure the RIS to maintain a high SNR at the reader while enhancing her own signal reception.
This limitation may result in the reader experiencing a noticeable degradation in SNR, thus, her ability to remain covert is compromised.
Despite this, the secrecy rate will still be negatively impacted due to Eve’s partial success in capturing the signal.

\subsubsection{PLA Performance Diminution}

The reliability of PLA in RIS-aided MBC systems are critically dependent on the secure and accurate functioning of RIS. When Eve compromises the RIS, she introduces several challenges that can degrade PLA performance by enhanced jamming capabilities, a reduction in secrecy rates, or an increase in false negatives.
In the context of a compromised RIS jamming, Eve has the ability to manipulate the RIS phase shifts to cause intentional destructive interference patterns at the reader, degrading the SNR at the reader, directly impacting the reliability of the signal detection and authentication process. 
The capability to perform such attacks without significant power expenditure (green jamming) makes it particularly stealthy and challenging to detect and counteract.
In addition, a compromised RIS can significantly diminish the secrecy rate, which is a measure of the confidentiality of the communication between the tag and the reader. As discussed, Eve can enhance her SNR by optimally manipulating the RIS, leading to an increased data rate at Eve’s end while potentially maintaining or slightly decreasing the data rate at the reader, thereby reducing the secrecy rate calculated as \eqref{SC1}. 
This breach in confidentiality can undermine the authenticity checks performed by the PLA, as Eve could potentially replicate or manipulate the communication data to impersonate the tag or the reader.
Furthermore, the manipulation of signal paths via a compromised RIS can lead to inconsistencies in the received signal characteristics at the reader compared to those expected based on legitimate channel conditions. Such inconsistencies can result in authentication errors, specifically increasing the likelihood of false negatives, where legitimate communications are incorrectly flagged as inauthentic (see Section \ref{Mal_RIS_SubSim} for more details). 

\section{Performance Evaluation} \label{sec_results}
Through simulations across various system settings and threat scenarios, we first evaluate the performance of the proposed PLA with a trusted RIS. Specifically, we examine the authentication accuracy, the impact of varying distances between the tag and reader, and the effect of different numbers of RIS elements on the overall system performance. Additionally, we assess the destructive impact of a compromised RIS on the secrecy capacity and PLA performance within MBC systems. This includes analyzing how unauthorized control over RIS affects PLA reliability and security, and evaluating the average secrecy capacity (ASC) under diverse channel conditions.

\subsection{Experimental Setting}

\subsubsection{Simulation Setup}
Our simulation setup meticulously recreates realistic communication environments, where a tag, a reader, and a RIS are subjected to different attack vectors. We examine three distinct threat scenarios in our simulation setting: a tag under attack, a reader under attack, and an RIS under attack. In the first two scenarios, the attacker impersonates a legitimate tag or reader using intercepted ID information to inject unauthorized messages. In the latter scenario, we evaluate the impact on the overall system’s security when the intelligent surface itself is compromised.
We assume an indoor RIS-aided MBC environment like a smart home where the tag, reader, RIS, and Eve are located in fixed positions. 
\begin{figure} [t]
    \centering
\includegraphics[width=0.40\textwidth]{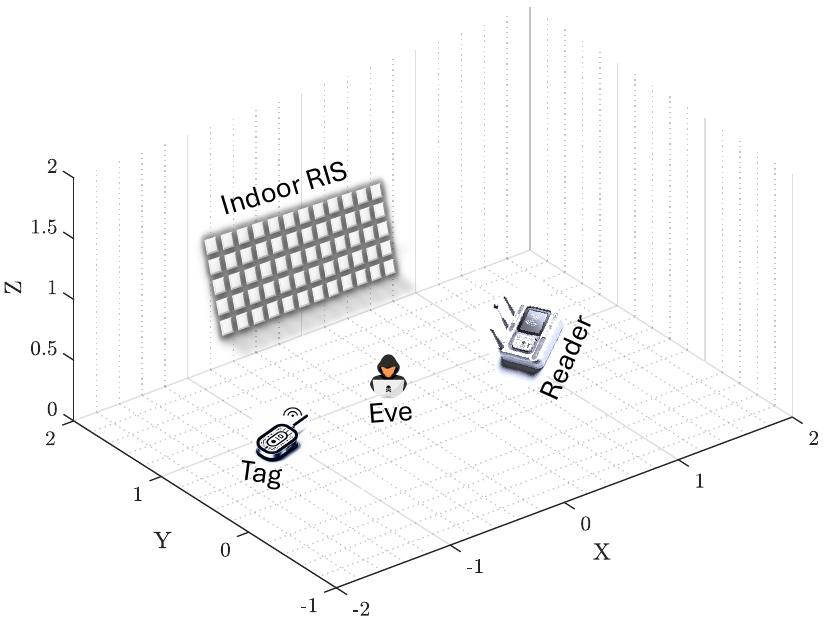}
\vspace{-10pt}
    \caption{Simulation setup.}
   \vspace{-10pt}
 \label{simsetup}
\end{figure}
Fig. \ref{simsetup} shows an illustration of our simulation setup. The RIS is strategically positioned to boost the received signal power at the tag and the reader (when is trusted). Eve is situated between the tag and the reader to fulfill her malicious purposes. 
This configuration allows for a comprehensive PLA performance analysis with fixed distances as $d_{R\Theta}= d_{T\Theta}= d_{TE}= d_{R E}= 1$m and having Eve a bit closer to the RIS (for the worst case scenario) $d_{E \Theta }=80$cm, suitable for indoor MBC use cases. 
Additional simulation parameters include a spectral efficiency of $R_s=1$ bps/Hz, noise power at the reader and Eve of $\sigma^2_R=-30$ dBm and $\sigma^2_E=-20$ dBm, respectively, a maximum average source power of $P_s=1$ dBm, and a path loss exponent of $\chi_1 =3.5$ and $\chi_2 =2.5$ for the direct and RIS-aided links, respectively.

\subsubsection{Performance Metrics}
To quantify the performance of our authentication scheme, we consider several metrics with a primary focus on the authentication accuracy. This metric provides insight into the ability of the system to accurately identify and validate legitimate endpoints while thwarting unauthorized access attempts. Following authentication processes at both the tag and the reader, the authentication outcome can be categorized as either $Accept$ or $Reject$. To assess the performance of our PLA method, we employ the following evaluation metrics: 1) Authentication Rate, representing the true positive rate indicating the accurate acceptance of legitimate tag or reader by our scheme, and 2) False Acceptance Rate (FAR), signifying the false positive rate, which measures the rate at which Eve is incorrectly accepted by our scheme. Subsequently, we utilize a Receiver Operating Characteristic (ROC) curve to illustrate the balance between the authentication rate and FAR across various authentication thresholds, threat scenarios, and system configurations. 
We also employ another performance metric known as ASC to demonstrate the impact of a compromised RIS on the secrecy performance of RIS-aided MBC.
ASC denotes the mean value of the secrecy rate, depicted in \eqref{SC1}, under diverse channel conditions, playing a vital role in evaluating secrecy performance. Algorithm \ref{alg:compute-asc} illustrates the method for calculating ASC in this study.
It should be highlighted that due to the complexity of deriving tractable closed-form expressions for the probability density function (PDF) and cumulative distribution function (CDF) of the SNR at the reader from the received signal in \eqref{y_R_eq1}, it is not feasible to obtain concise analytical expressions for the ASC within the proposed system model. Consequently, to illustrate the behavior of ASC in the presence of both trusted and malicious RIS in our system model, we rely solely on Monte Carlo simulation approach.
\begin{algorithm}[t]
\caption{Compute ASC}
\label{alg:compute-asc}
\begin{algorithmic}[1]
\footnotesize
\State \textbf{Input:} $\gamma_R$, $\gamma_E$
\State \textbf{Initialize:} $\bar{\gamma}_{R}$, $N_{sim}$ \Comment{Average SNR at the reader and number of simulation samples, respectively}
\State \textbf{Output:} $\bar{C}_s$ containing the ASC
\State $C_s \gets \text{zeros}(\text{length}(\bar{\gamma}_{R}), N_{sim})$ \Comment{Secrecy rate}
\For{$k = 1$ \textbf{to} \text{length}($\bar{\gamma}_{R}$)}
    \State $C_s[k,:] \gets (\log_2(1+\gamma_R) - \log_2(1+\gamma_E)) \cdot (\gamma_R > \gamma_E)$ 
\EndFor
\State $\bar{C}_s \gets \text{mean}(C_s, 2)$ \Comment{Average secrecy capacity}
\end{algorithmic}
\end{algorithm}

\subsection{Simulation Results and Discussion}
Here, we further discuss the performance of the proposed PLA in various system configurations and across all mentioned attack scenarios.

\subsubsection{Tag Is Under Attack} 
\begin{figure*}[t]
\centering
    \begin{minipage}[b]{0.30\linewidth}
        \centering
        \raggedleft
        \includegraphics[width=\linewidth]{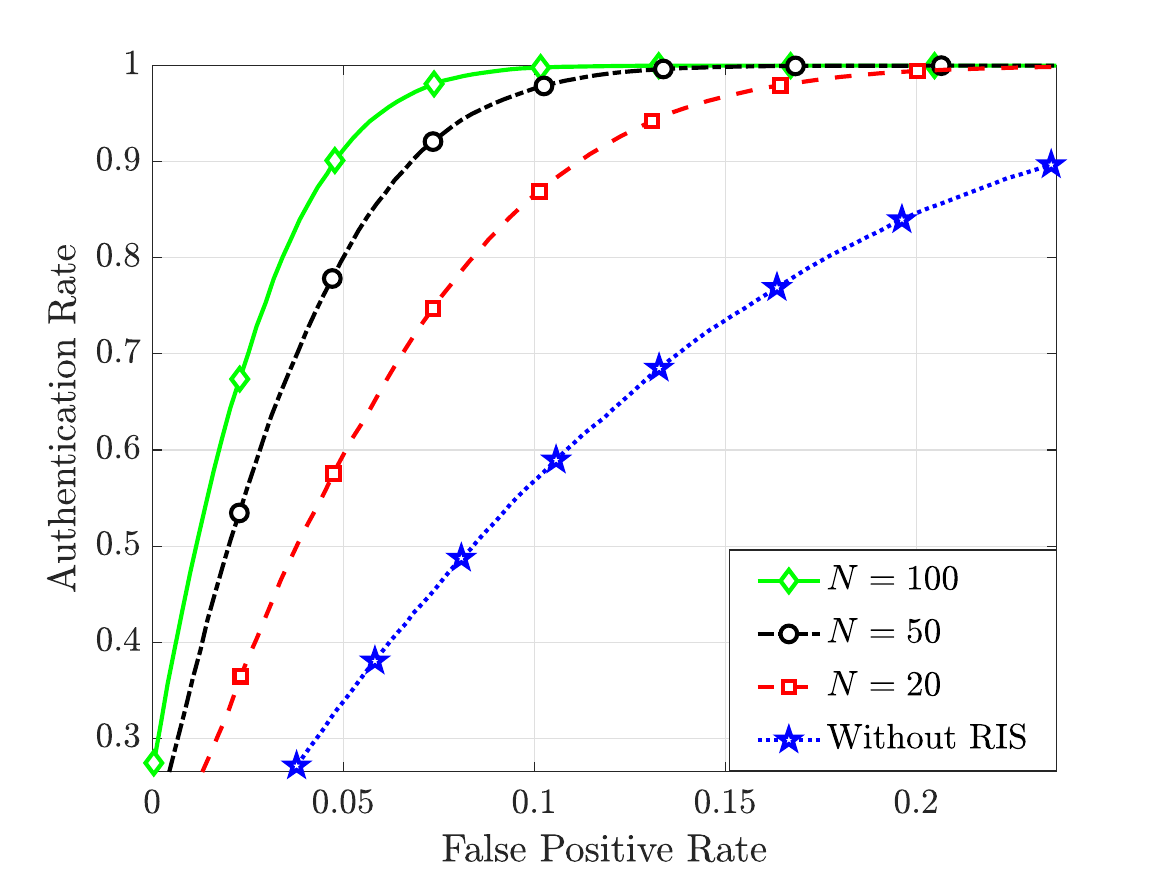}
        \vspace{-12pt}
    \caption{ROC for the tag authenticating the reader with different numbers of RIS elements.}
    \vspace{-5pt}
    \label{ROC_tag_N}
    \end{minipage}
    \hfill
    \begin{minipage}[b]{0.30\linewidth}
        \centering
       \includegraphics[width=\linewidth]{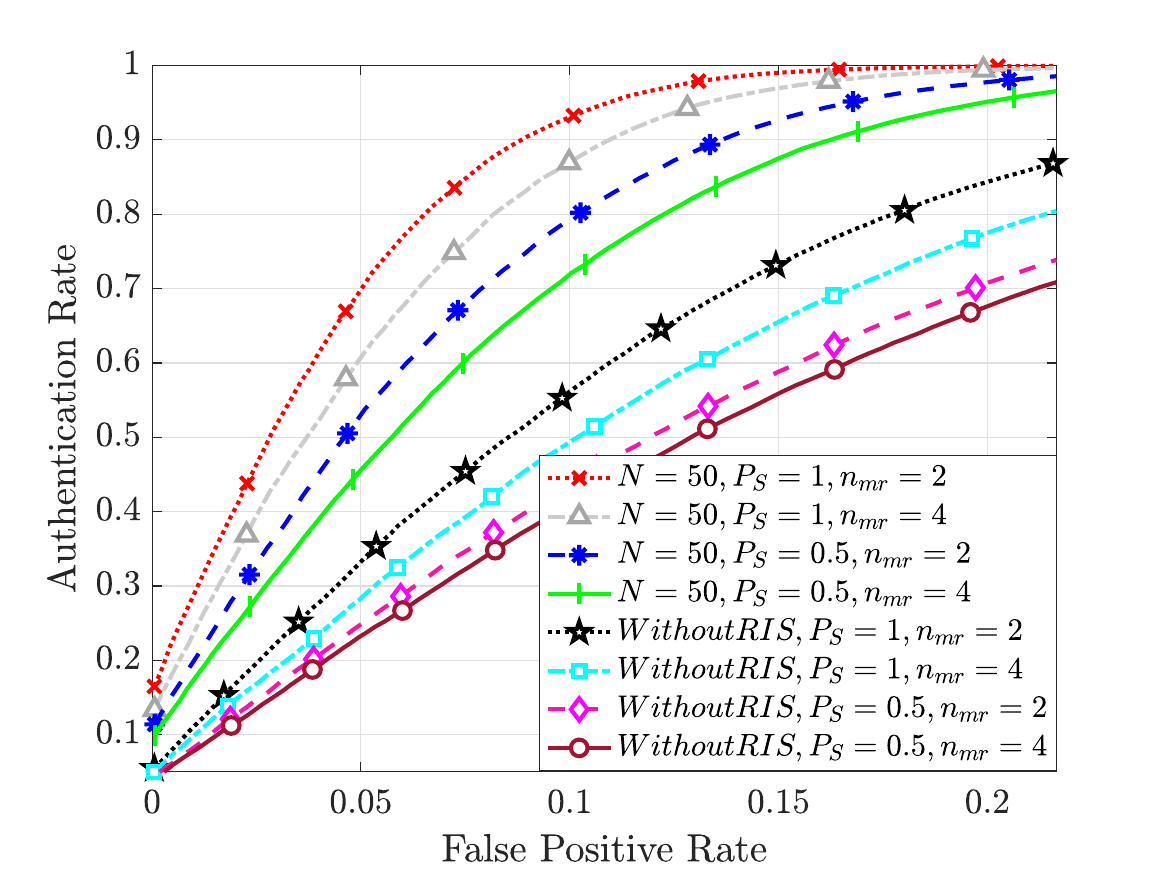}
       \vspace{-12pt}
    \caption{Authentication rate vs. $n_{mr}$ and $P_S$ for different numbers of RIS elements.}
    \vspace{-5pt}
    \label{ROC_tag_P}
   \end{minipage}
    \hfill
    \begin{minipage}[b]{0.30\linewidth}
       \raggedright
       \includegraphics[width=\linewidth]{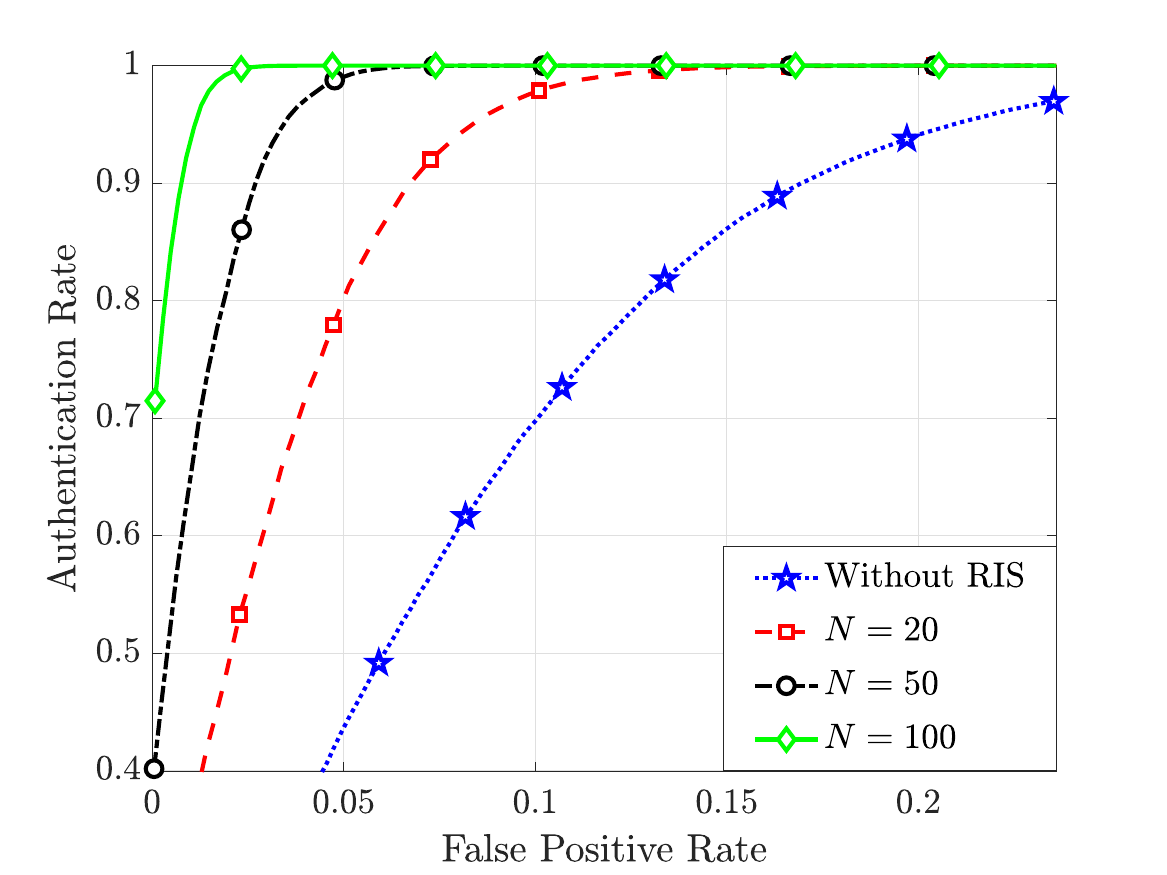}
       \vspace{-12pt}
    \caption{ROC for the reader authenticating the tag with different numbers of RIS elements.}
    \vspace{-5pt}
    \label{ROC_R_RIS_N}
    \end{minipage}
\end{figure*}

Fig. \ref{ROC_tag_N} illustrates the ROC curves for the tag authenticating the reader with different numbers of RIS elements. It is obvious from the figure that the presence of RIS significantly improves the authentication performance compared to the scenario without RIS. Furthermore, as the number of RIS elements increases, the authentication rate increases for any given false positive rate. Specifically, with \(N = 100\), the authentication rate approaches $1$ with a very low false positive rat, indicating highly reliable authentication. This improvement is attributed to the stable power delivery at the tag's energy detector due to the optimal phase configuration of the RIS elements, resulting in a more reliable output voltage profile, thereby enhancing the authentication performance at the tag's end.
Fig. \ref{ROC_tag_P} presents the PLA performance under varying conditions of the density of the malicious readers ($n_{mr}$) and the source power $P_S$ with different numbers of RIS elements. It is evident that using RIS significantly boosts the authentication performance compared to scenarios without RIS. The graph also shows that increased power $P_S$ results in better authentication performance. The variation in the density of attackers also impacts the authentication rate, where higher densities yield worse PLA performance.

\subsubsection{Reader Is Under Attack} 

Fig. \ref{ROC_R_RIS_N} presents the ROC curves for the reader authenticating the tag with different numbers of RIS elements. The figure clearly demonstrates that the presence of RIS significantly enhances authentication performance compared to the scenario without RIS. As the number of RIS elements increases, the authentication rate improves for any given false positive rate. 
This improvement can be attributed to several key factors: The RIS optimally configures the phase of each of its elements to focus the reflected signals towards the reader. This optimal phase configuration maximizes the constructive interference of the backscattered signals from the tag, significantly enhancing RSS at the reader. Higher RSS also leads to more distinct and reliable signal features, making it easier for the reader to authenticate the tag accurately. 
With an increased number of RIS elements, there are more degrees of freedom for configuring the reflection phase shifts. This allows for finer control over the signal environment, enabling the RIS to better compensate for multipath fading and other channel impairments. Consequently, the reader can receive a more coherent and strong signal from the tag, enhancing the reliability of the authentication process.
%
%
Table \ref{Table_Compare_Auth} examines the impact of varying distances between the tag and the reader on the PLA performance in the absence of any attackers. The results illustrate that as the distance increases, the authentication rate tends to decrease, particularly without RIS assistance. However, with the integration of RIS, the authentication performance remains relatively stable even at larger distances, with a rate of $98.85$\% observed at $\text{d}_{TR}=6$m for $N=100$. 
We also compare the performance of the proposed RIS-aided PLA scheme with the recently published related work, BCAuth \cite{BCAuth}, which is considered the most efficient for MBC systems. Compared to the scenario with $N=100$, our scheme outperforms BCAuth \cite{BCAuth}, especially at longer distances. Specifically, when we increase $\text{d}_{TR}$ from 2m to 6m, our scheme shows a performance degradation of only $0.94$\%, whereas BCAuth \cite{BCAuth} exhibits a degradation of $5.06$\%.
It is also worth mentioning that BCAuth \cite{BCAuth} performs more complex signal analysis at the reader’s end, including computing RSS, angle of arrival (AoA), and cluster-based authentication. In contrast, our scheme relies solely on computing RSS. This demonstrates the simpler design of our reader while achieving better performance in both accuracy and secure coverage area. One can also observe the impact of having more complex signal processing power at the reader on the PLA performance by comparing the ``No RIS'' scenario of our scheme with BCAuth \cite{BCAuth}.
Thus, these findings underscore the importance of RIS in extending the secure coverage range where consistent PLA performance can be maintained, which is crucial for BC systems in which the transmission range is limited and the distance between the tag and the reader profoundly impacts PLA performance. 
Fig. \ref{ROC_RIS_Eve_num} depicts the PLA performance with varying numbers the malicious tags. Despite the presence of attackers results in degrading the PLA performance, the ROC curves demonstrate that the authentication performance remains acceptable, especially with a higher number of RIS elements. The RIS effectively mitigates the impact of attackers by enhancing the RSS and providing more distinct signal features for authentication, thereby ensuring robust performance even in the presence of more adversaries in the network.
\begin{table}[t]
\centering
\caption{Authentication Performance Comparison for Different Tag-to-Reader Distances}
\begin{tabular}{cc|c|c|c|c|c|}
\cline{3-7}
\multicolumn{2}{c|}{}   & $2$m     & $3$m     & $4$m     & $5$m     & $6$m     \\ \hline
\multicolumn{2}{|c|}{BCAuth \cite{BCAuth}   } & $96.88$\% & $95.00$\% & $93.48$\% & $92.62$\% & $91.98$\% \\ \hline
\multicolumn{1}{|c|}{\multirow{4}{*}{\rotatebox[origin=c]{90}{Ours}}} & No RIS & $92.87$\% & $90.51$\% & $88.68$\% & $87.93$\% & $87.44$\% \\ \cline{2-7} 
\multicolumn{1}{|c|}{}  & $N=20$   & $98.23$\% & $97.67$\% & $96.43$\% & $95.31$\% & $94.83$\% \\ \cline{2-7} 
\multicolumn{1}{|c|}{}  & $N=50$   & $99.20$\% & $98.85$\% & $98.61$\% & $97.59$\% & $96.98$\% \\ \cline{2-7} 
\multicolumn{1}{|c|}{}  & $N=100$  & $99.79$\% & $99.58$\% & $99.39$\% & $99.10$\% & $98.85$\% \\ \hline
\end{tabular}
\label{Table_Compare_Auth}
\end{table}

\subsubsection{RIS Is Under Attack} \label{Mal_RIS_SubSim} 

\begin{figure*}[t]
\centering
    \begin{minipage}[b]{0.30\linewidth}
        \centering
        \raggedleft
        \includegraphics[width=\linewidth]{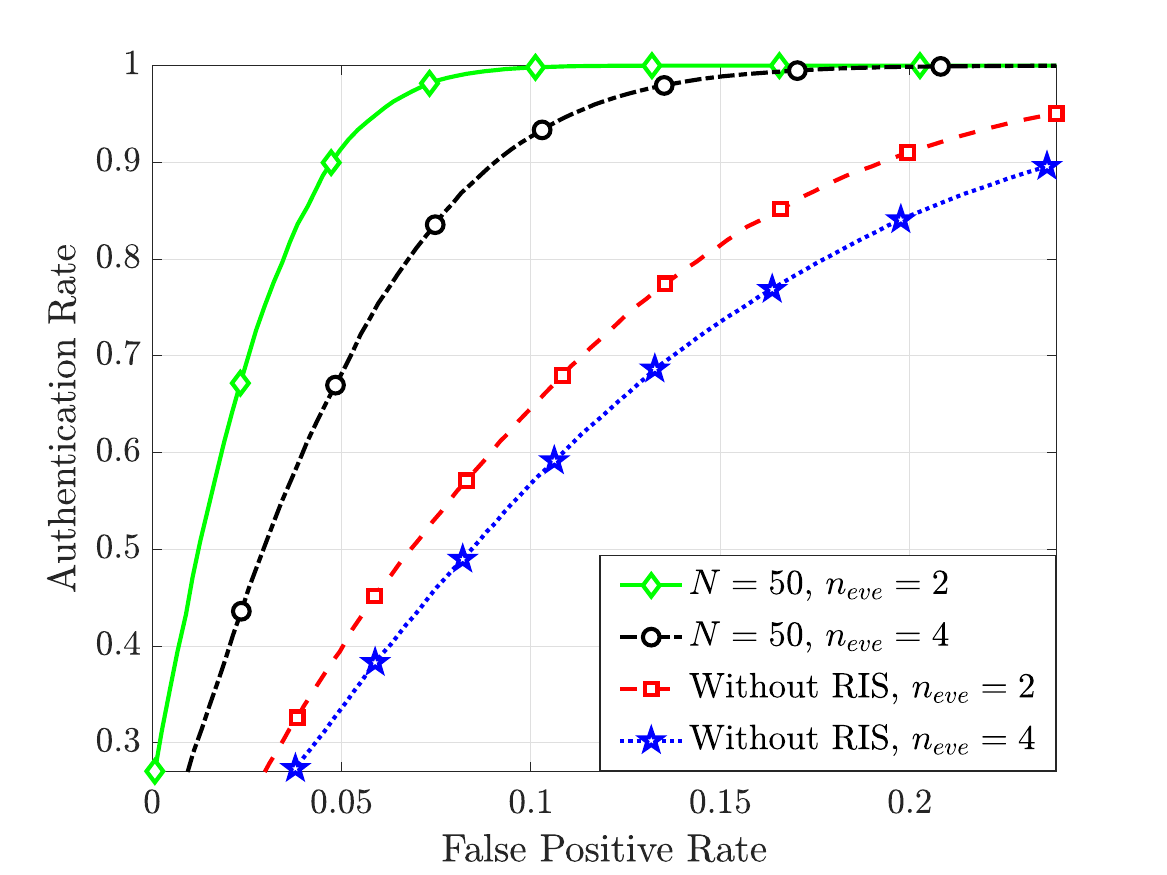}
        \vspace{-12pt}
    \caption{Authentication rate vs. the density of attackers for different numbers of RIS elements.}
    \vspace{-4pt}
    \label{ROC_RIS_Eve_num}
    \end{minipage}
    \hfill
    \begin{minipage}[b]{0.30\linewidth}
        \centering
       \includegraphics[width=\linewidth]{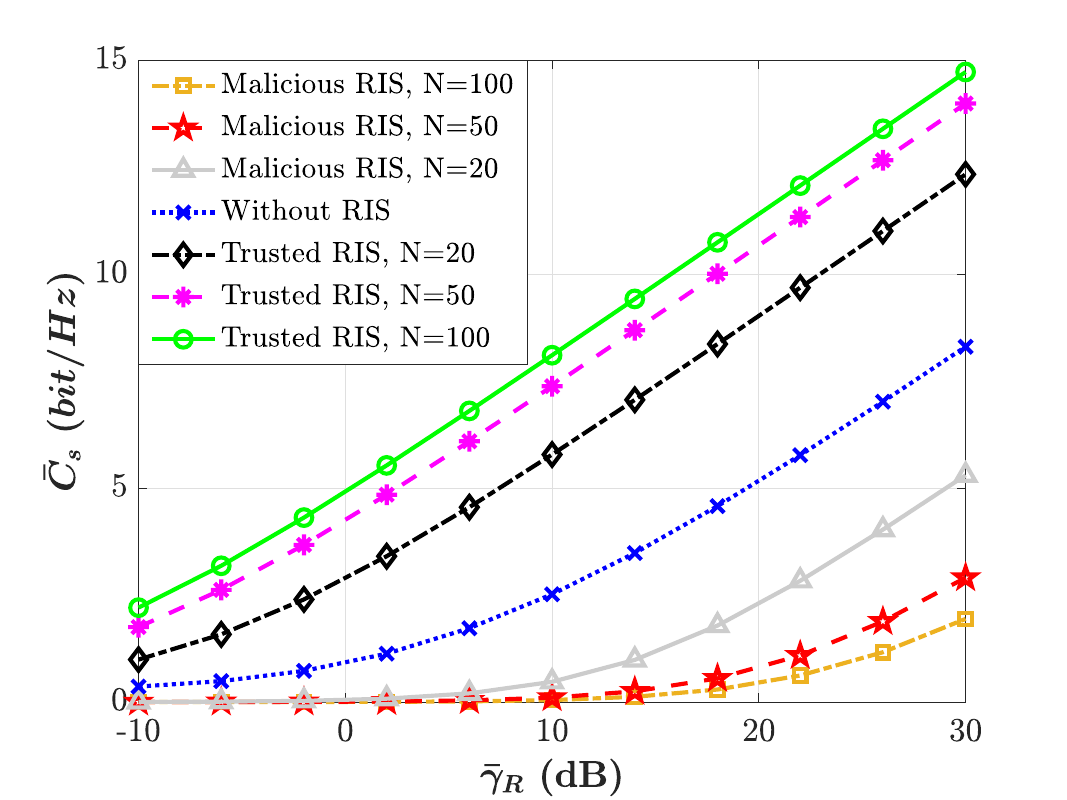}
       \vspace{-12pt}
    \caption{ASC versus $\Bar{\gamma}_{R}$ in the presence of both trusted and malicious RISs.}
        \vspace{-4pt}
    \label{Mal_RIS_SR}
   \end{minipage}
    \hfill
    \begin{minipage}[b]{0.30\linewidth}
       \raggedright
       \includegraphics[width=\linewidth]{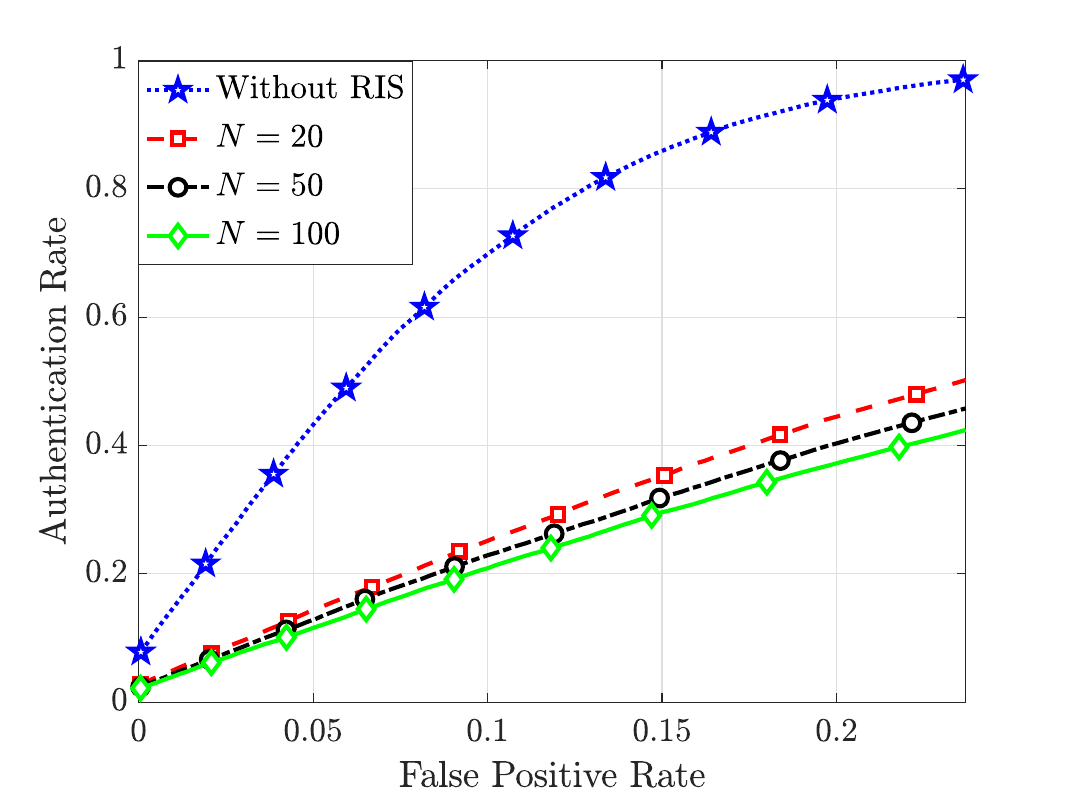}
       \vspace{-12pt}
    \caption{ROC for the reader authenticating the tag in the presence of a malicious RIS.}
    \vspace{-4pt}
    \label{Mal_RIS_ROC}
    \end{minipage}
\end{figure*}

Fig. \ref{Mal_RIS_SR} illustrates the ACS (\( \tilde{C}_s \)) as a function of the average received SNR at the reader (\( \bar{\gamma}_R \)) for scenarios with trusted and malicious ISs across varying numbers of RIS elements. The figure delineates that trusted RIS configurations substantially enhance secrecy capacity, with improvements scaling with the number of RIS elements. For instance, with $N=100$, the system shows the highest secrecy improvements, confirming that more RIS elements can more effectively shape and direct signals to bolster secure communications. Conversely, the presence of a malicious RIS drastically reduces the secrecy performance, particularly as the number of elements increases, demonstrating how a compromised RIS can be exploited to severely disrupt communication security. 
The comparison with the scenario without RIS showcases that while a non-RIS environment offers moderate secrecy improvements with increasing \( \bar{\gamma}_R \), it is outperformed by scenarios involving a trusted RIS and outperforms those with a malicious RIS. These observations underscore the significant impact of RIS configurations on communication security in MBC systems, highlighting the dual potential of RIS technology to either enhance or compromise the secrecy performance depending on its operational integrity.
Fig.~\ref{Mal_RIS_ROC} demonstrates that the authentication performance significantly drops in the presence of malicious RIS compared to the scenario without RIS. The curve for $N=20$ elements shows a noticeable reduction in authentication rate, which further deteriorates as the number of RIS elements increases to $N=50$ and $N=100$. This decline in performance is attributed to the increased ability of the adversary to manipulate the phase shifts and signal paths, causing greater false negative during the authentication process at the reader, hindering the reader's ability to correctly authenticate the tag. Additionally, the higher the number of compromised RIS elements, the more control the attacker has over the signal, further degrading the system's secrecy rate and making it difficult for the reader to distinguish between legitimate and malicious signals.

\section{Conclusion 
}
\label{sec_conclusion}
This paper introduced a RIS-aided PLA scheme for MBC systems, which primarily leveraged the unique effects of RIS on BC physical layer attributes to enhance PLA performance in the studied system model. Additionally, by exploiting the simple construction of energy detector circuits in tags and the optimal power delivery of RIS in BC, we proposed a lightweight method for resource-limited passive tags to authenticate the reader, thereby addressing the lack of mutual authentication in BC systems.
Extensive simulations under various system settings demonstrated that integrating RIS with MBC systems not only improves authentication performance but also extends the operational PLA coverage range, ensuring robust performance even over greater distances between the tag and the reader.
Furthermore, we conducted a comprehensive security analysis of potential risks under three different threat scenarios. Our findings indicated that while a trusted RIS significantly enhances PLA performance and secures the system against various types of attacks, a malicious RIS can severely degrade both authentication and secrecy performance. This dual-edged impact of RIS highlights the critical need for secure control over RIS elements to fully harness their potential in enhancing MBC system security.



\begin{thebibliography}{1}




\bibitem{BC_Battery_CST2023}
T. Jiang, et al. ``Backscatter Communication Meets Practical Battery-Free Internet of Things: A Survey and Outlook," \textit{IEEE Communications Surveys \& Tutorials}, vol. 25, no. 3, pp. 2021--2051, 2023.

\bibitem{BC_survey_01}
C. Yao, Y. Liu, X. Wei, G. Wang, and F. Gao, ``Backscatter Technologies and the Future of Internet of Things: Challenges and Opportunities,'' \textit{Intelligent and Converged Networks}, vol. 1, no. 2, pp.170--180, 2020. 



\bibitem{RIS_BC_frontier_6G}
S. Basharat, et al.,
``Reconfigurable intelligent surface-assisted backscatter communication: A new frontier for enabling 6G IoT networks,'' \textit{IEEE Wireless Communications}, vol. 29, no. 6, pp. 96--103, 2022.

\bibitem{RIS_BC_Survey_Proceeding}
Y. C. Liang, et al. ``Backscatter communication assisted by reconfigurable intelligent surfaces." \textit{Proceedings of the IEEE}, vol. 110, no. 9, pp. 1339--1357, 2022.





\bibitem{ref18}
R. Fara, et al, ``A prototype of reconfigurable intelligent surface with continuous control of the reflection phase,'' \textit{IEEE Wireless Communications}, vol. 29, no. 1, pp. 70--77, 2022.



\bibitem{ref20}
X. Jia, et al, ``Intelligent reflecting surface-assisted bistatic backscatter networks: Joint beamforming and reflection design,'' \textit{
IEEE Trans. Green Commun. Netw.}, vol. 6, no. 2, pp. 799--814, 2021.





\bibitem{ref26}
A. Hakimi, et al, ``Sum Rate Maximization of MIMO Monostatic Backscatter Networks by Suppressing Residual Self-Interference,'' \textit{IEEE Transactions on Communications}, vol. 71, no. 1, pp. 512--526 2022.











\bibitem{ref36}
Z. Wang, 
et al, ``Deep Unfolding-Based Joint Beamforming and Detection Design for Ambient Backscatter Communications with IRS,'' \textit{IEEE Communications Letters}, vol. 27, no. 4, pp. 1145--1149, 2023.





\bibitem{ref29}
S. Zargari, et al, ``Energy-Efficient Hybrid Offloading for Backscatter-Assisted Wirelessly Powered MEC with Reconfigurable Intelligent Surfaces,'' \textit{IEEE Trans. Mobile Comput.}, vol. 22, no. 9, pp. 5262--5279, 2022.







%




\bibitem{ref3941}
D. Galappaththige, F. Rezaei, C. Tellambura, and S. Herath, ``Optimizing Passive Tag Performance with Reconfigurable Intelligent Surfaces in Bistatic Backscatter Networks,'' \textit{IEEE Transactions on Vehicular Technology}, 2024. 





\bibitem{PLA_survey_01}
N. Xie, Z. Li, and H. Tan, ``A survey of physical-layer authentication in wireless communications,'' \textit{IEEE Communications Surveys \& Tutorials}, vol. 2, no. 1, pp. 282--310, 2020. 



\bibitem{RSS_Auth_01}
Y. Sheng, 
et al, “Detecting 802.11 mac layer spoofing using received signal strength,” \textit{in IEEE 27th Conference on Computer Communications (INFOCOMM)}, pp. 1768--1776, 2008.



\bibitem{APAuth}
J. D. Chang, J. J. Li, Y. S. Yang, Y. F. Zhang, M. Kaveh, and Z. Yan, ``APAuth: Authenticate an Access Point by Backscatter Devices”, \textit{IEEE 2024 IEEE International Conference on Communications (ICC)}, 
2024. 










\bibitem{RFID_Auth_07}
P. D'Arco and A. De Santis, ``On ultralightweight RFID authentication protocols,'' \textit{IEEE Transactions on Dependable and Secure Computing}, vol. 8, no. 4, pp. 548--563, 2010. 





\bibitem{Server_indp_RFID}
B. Wang and M. Ma, ``A server independent authentication scheme for RFID systems,’’ \textit{IEEE Transactions on Industrial Informatics}, vol. 8, no. 3, pp. 689--696, 2012.

\bibitem{USI_RFID_Auth}
L. Gao, L. Zhang, F. Lin, and M. Ma, ``Secure RFID authentication schemes based on security analysis and improvements of the USI protocol,’’ \textit{IEEE Access}, vol. 7, pp. 8376--8384, 2019.

\bibitem{RFID_Auth_Analyst_01}
 M. Hosseinzadeh, et al, ``An enhanced authentication protocol for RFID systems,'' \textit{IEEE Access}, vol. 8, pp. 126977--126987, 2020.



\bibitem{RFID_Auth_ECC_01}
N. Dinarvand and H. Barati, ``An efficient and secure RFID authentication protocol using elliptic curve cryptography,'' \textit{Wireless Networks}, vol. 25, no. 1, pp. 415--428, 2019.

\bibitem{RFID_Auth_ECC_02}
S. Izza, M. Benssalah, and K. Drouiche, ``An enhanced scalable and secure RFID authentication protocol for WBAN within an IoT environment,'' \textit{Journal of Information Security and Applications}, vol. 58, p. 102705, 2021.

\bibitem{RFID_Auth_ECC_03}
U. Ali, et al, ``RFID authentication scheme based on hyperelliptic curve signcryption,'' \textit{IEEE Access}, vol. 9, pp. 49942--49959, 2021.



\bibitem{RFID_Auth_PUF_01}
T. F. Lee, K. Lin, Y. Hsieh, and K. Lee, ``Lightweight cloud computing-based RFID authentication protocols using PUF for e-healthcare systems,'' \textit{IEEE Sensors Journal}, vol. 23, no. 6, pp. 6338--6349, 2023.

\bibitem{RFID_Auth_PUF_02}
P. Gope, J. Lee, and T. Q. Quek, ``Lightweight and practical anonymous authentication protocol for RFID systems using physically unclonable functions,'' \textit{IEEE Transactions on Information Forensics and Security}, vol. 13, no. 11, pp. 2831--2843, 2018.




\bibitem{BC_PLA_01}
D. Zanetti, B. Danev, and S. Apkun, “Physical-layer identification of UHF RFID tags,” \textit{in Proc. 16th Annu. Int. Conf. Mobile Comput. Netw. (MobiCom)}, pp. 353--364, 2010.

\bibitem{BC_PLA_02}
J. Han, et al., ``GenePrint: Generic and accurate physical-layer identification for UHF RFID tags,” \textit{IEEE/ACM Trans. Netw.}, vol. 24, no. 2, pp. 846--858, 2015. 

\bibitem{BC_PLA_03}
A. Mehmood, W. Aman, M. M. U. Rahman, M. A. Imran, and Q. H. Abbasi, ``Preventing identity attacks in RFID backscatter communication systems: A physical-layer approach,” \textit{in Proc. Int. Conf. U.K. China Emerg. Technol. (UCET)}, pp. 1--5, 2020.

\bibitem{BC_PLA_04}
G. Wang et al., ``Towards replay-resilient RFID authentication,” \textit{in Proc. 24th Annu. Int. Conf. Mobile Comput. Netw.}, pp. 385--399, 2018.

\bibitem{BC_PLA_05}
B. Danev, T. S. Benjamin, and S. Capkun, “Physical-layer identification of RFID devices,” \textit{in Proc. USENIX Secur. Symp.}, pp. 199--214, 2009.

\bibitem{BC_PLA_06}
Z. Luo, W. Wang, Q. Huang, T. Jiang, and Q. Zhang,  ``Securing IoT devices by exploiting backscatter propagation signatures,'' \textit{IEEE Transactions on Mobile Computing}, vol. 21, no. 12, pp. 4595--4608, 2021.

\bibitem{BCAuth}
P. Wang, Z. Yan, and K. Zeng, ``Bcauth: Physical layer enhanced authentication and attack tracing for backscatter communications,” \textit{IEEE Transactions on Information Forensics and Security}, vol. 17, pp. 2818--
2834, 2022.

\bibitem{PLA_AmBC_NOMA}
X. Li, et al., ``Physical-layer authentication for ambient backscatter-aided NOMA symbiotic systems,” \textit{IEEE Transactions on Communications}, vol. 71,
no. 4, pp. 2288--2303, 2023.

\bibitem{BatAu}
Y. Yang, et al.,
``BatAu: A Batch Authentication Scheme for Backscatter Devices in a Smart Home Network,'' \textit{In ICC 2023-IEEE International Conference on Communications}, pp. 4528--4533, 2023.

\bibitem{Cross-Domain-PLA}
G. Zhang, Q. Hu, Y. Zhang, Y. Dai, and T. Jiang,  ``Lightweight Cross-Domain Authentication Scheme for Securing Wireless IoT Devices Using Backscatter Communication,'' \textit{IEEE Internet of Things Journal}, 2024.


\bibitem{RIS-PLA_01}
N. Gao, et al., ``RIS-Assisted Physical Layer Authentication for 6G Endogenous Security,'' \textit{arXiv preprint arXiv:2309.07736}, 2023. 

\bibitem{RIS-PLA_02}
P. Zhang, Y. Teng, Y. Shen, X. Jiang, and F. Xiao, ``Tag-based PHY-layer authentication for RIS-assisted communication systems,'' \textit{IEEE Transactions on Dependable and Secure Computing}, 2023.

\bibitem{RIS-PLA_03}
M. M. Selim, and S. Tomasin, ``Physical Layer Authentication With Simultaneous Reflecting and Sensing RIS,'' \textit{In 2023 IEEE 97th Vehicular Technology Conference (VTC2023-Spring)} pp. 1--5, 2023.

\bibitem{RIS-PLA_04}
J. He, M. Niu, P. Zhang, and C. Qin, ``Enhancing PHY-Layer Authentication in RIS-Assisted IoT Systems With Cascaded Channel Features,'' \textit{IEEE Internet of Things Journal}, 2024.

\bibitem{RIS-PLA_05}
M.A. Shawky, et al. ``Reconfigurable Intelligent Surface-Assisted Cross-Layer Authentication for Secure and Efficient Vehicular Communications,'' \textit{arXiv preprint arXiv:2303.08911}, 2023. 






\bibitem{BD_Eng_detector1}
R., Reed, F.L. Pour, and D.S., Ha, ``An energy efficient RF backscatter modulator for IoT applications,'' \textit{In 2021 IEEE International Symposium on Circuits and Systems (ISCAS)}, pp. 1--5, 2021.





\bibitem {BjornsonCSI}
E. Björnson, et al.,
``Reconfigurable intelligent surfaces: A signal processing perspective with wireless applications,” \textit{IEEE Signal Process. Mag.}, vol. 39, no. 2, pp. 135–158, 2022.


\bibitem{ref57}
Y. Liu, Y. Ye, and R. Q. Hu  ``Secrecy outage probability in backscatter communication systems with tag selection,'' \textit{IEEE Wireless Communications Letters}, vol. 10, no. 10, pp. 2190--2194, 2021.



\bibitem{Indoor_RIS_01}
S. Kayraklik, et al. ``Indoor Measurements for RIS-Aided Communication: Practical Phase Shift Optimization, Coverage Enhancement, and Physical Layer Security,'' \textit{IEEE Open Journal of the Communications Society}, 2024. 

\bibitem{Indoor_RIS_02}
J. Yuan, O. Franek, H. Fang, and P. Popovski, ``Indoor RIS-Assisted Wireless System with Location-Based Reflective Patterns,'' \textit{IEEE Transactions on Communications}, 2024.





\bibitem{metasurfaceRISmanipu6G}
H., Alakoca, et al., ``Metasurface manipulation attacks: Potential security threats of RIS-aided 6G communications,'' \textit{IEEE Communications Magazine}, vol. 61, no. 1, pp. 24--30, 2022. 

\bibitem{RIS_attack01}
H. Chen and Y. Ghasempour, ``Malicious mmWave reconfigurable surface: Eavesdropping through harmonic steering,'' \textit{In Proceedings of the 23rd Annual International Workshop on Mobile Computing Systems and Applications}, pp. 54--60, 2022.

\bibitem{RIS_attack02}
Z. Wei, B. Li and W. Guo, ``Adversarial reconfigurable intelligent surface against physical layer key generation,'' \textit{IEEE Transactions on Information Forensics and Security}, vol. 18, pp. 2368--2379, 2023. 

\bibitem{RIS_attack03}
L. Hu, G. Li, H. Luo, and A. Hu, ``On the RIS manipulating attack and its countermeasures in physical-layer key generation,'' \textit{In 2021 IEEE 94th Vehicular Technology Conference (VTC2021-Fall)}, pp. 1--5, 2021. 

\bibitem{RIS_attack05}
Y. Wang, H. Lu, D. Zhao, Y. Deng, and A. Nallanathan, ``Wireless communication in the presence of illegal reconfigurable intelligent surface: Signal leakage and interference attack,'' \textit{IEEE Wireless Communications}, vol. 29, no. 3, pp. 131--138, 2022. 

\bibitem{RIS_attack06}
F. Naeem, M. Ali, G. Kaddoum, C. Huang, and C. Yuen, ``Security and privacy for reconfigurable intelligent surface in 6G: A review of prospective applications and challenges,'' \textit{IEEE Open Journal of the Communications Society}, 2023. 





\bibitem{Tag-Comparator01}
Y., Karimi, A., Athalye, S.R., Das, P.M. Djurić, and  M., Stanaćević, ``Design of a backscatter-based tag-to-tag system,'' \textit{In 2017 IEEE International Conference on RFID}, pp. 6--12, 2017.


\bibitem{RIS_MBC_01}
S. Idrees, S. Durrani, Z. Xu, X. Jia, and X. Zhou, ``Joint Active and Passive Beamforming for IRS-assisted Monostatic Backscatter Systems: An Unsupervised Learning Approach,'' \textit{IEEE Transactions on Machine Learning in Communications and Networking}, 2024. 




\bibitem{RISAntiJamm1}
X. Yuan, S. Hu, W. Ni, R. P. Liu, and X. Wang,``Joint user, channel, modulation-coding selection, and RIS configuration for jamming resistance in multiuser OFDMA systems,'' \textit{IEEE Transactions on Communications} vol. 71, no. 3, pp. 1631--1645, 2023.

\bibitem{RISAntiJamm2}
Y. Sun, et al., ``RIS-assisted robust hybrid beamforming against simultaneous jamming and eavesdropping attacks,'' \textit{IEEE Transactions on Wireless Communications }vol. 21, no. 11, pp. 9212--9231, 2022.

\bibitem{RISAntiJamm3}
S. Lin, et al., ``Secure multicast communications via RIS against eavesdropping and jamming with imperfect CSI,'' \textit{IEEE Transactions on Vehicular Technology}, 2023.




\bibitem{RISJamm1}
A. S. de Sena, J. Kibiłda, N. H. Mahmood,  A. Gomes, and M. Latva-aho, ``Malicious RIS Versus Massive MIMO: Securing Multiple Access Against RIS-Based Jamming Attacks,'' \textit{IEEE Wireless Communications Letters}, vol. 13, no. 4, pp. 989--993, 2024. 

\bibitem{RISJamm2}
P. Mackensen, et al., ``Spatial-Domain Wireless Jamming with Reconfigurable Intelligent Surfaces,'' \textit{arXiv preprint arXiv:2402.13773}, 2024. 

\bibitem{RISJamm3}
B. Lyu, et al.,
``IRS-based wireless jamming attacks: When jammers can attack without power,'' \textit{IEEE Wireless Communications Letters}, vol. 9, no. 10, pp. 1663--1667, 2020. 

\bibitem{RISJamm4}
Z. Lin, et al, ``Pain without gain: Destructive beamforming from a malicious RIS perspective in IoT networks,'' \textit{IEEE Internet of Things Journal} vol. 11, no. 5, pp. 7619--7629, 2024.


\bibitem{RISJamm6}
H. Huang, Y. Zhang, H. Zhang, C. Zhang, and Z. Han,  ``Illegal intelligent reflecting surface based active channel aging: When jammer can attack without power and CSI,'' \textit{IEEE Transactions on Vehicular Technology}, vol. 72, no. 8, pp. 11018--11022, 2023.



\bibitem{Mal_RIS_Dest_Beam}
S. Rivetti, Ö. T. Demir, E. Björnson, and M. Skoglund,  ``Malicious Reconfigurable Intelligent Surfaces: How Impactful can Destructive Beamforming be?,'' \textit{IEEE Wireless Communications Letters}, 2024.










\bibitem{RIS_SG_PLS01}
M. Kaveh, Z. Yan and R. Jäntti, ``Secrecy performance analysis of RIS-aided smart grid communications,'' \textit{IEEE Transactions on Industrial Informatics}, vol. 20, no. 3, pp. 5415-5427, 2024.


\bibitem{Mal_RIS_sideband_Eve}
H. Chen and Y. Ghasempour, ``Malicious mmWave reconfigurable surface: Eavesdropping through harmonic steering,'' \textit{In Proceedings of the 23rd Annual International Workshop on Mobile Computing Systems and Applications}, pp. 54--60, 2022. 

\bibitem{Mal_RIS_mmWave_Eve}
H. Chen, H. Saeidi, S. Venkatesh, K. Sengupta, and Y. Ghasempour, ``Wavefront Manipulation Attack via Programmable mmWave Metasurfaces: from Theory to Experiments,'' \textit{In Proceedings of the 16th ACM Conference on Security and Privacy in Wireless and Mobile Networks}, pp. 317--328, 2023. 




\end{thebibliography}
\end{document}